\documentclass[letterpaper,11pt]{article}

\usepackage{amsthm,amsfonts,amsmath}
\usepackage{fullpage}
\usepackage{itemsep}
\usepackage{hyperref}
\usepackage{graphicx}

\newtheorem{lemma}{Lemma}
\newtheorem{theorem}{Theorem}
\newtheorem{corollary}{Corollary}
\newtheorem{definition}{Definition}
\newtheorem{proposition}{Proposition}

\date{}

\newsavebox{\fmbox}
\newenvironment{fmpage}[1]
     {\medskip\begin{lrbox}{\fmbox}\begin{minipage}{#1}}
     {\end{minipage}\end{lrbox}\fbox{\usebox{\fmbox}}\medskip}

\newcommand{\ket}[1]{\lvert #1 \rangle}
\newcommand{\bra}[1]{\langle #1 \rvert} 
\newcommand{\norm}[1]{\lVert #1 \rVert} 
\newcommand{\abs}[1]{\lvert #1 \rvert} 
\def\size{\abs}
\newcommand{\ketbra}[1]{\lvert #1 \rangle\!\langle #1 \rvert} 
\newcommand{\ketbraa}[2]{\lvert #1 \rangle\!\langle #2 \rvert} 
\newcommand{\braket}[2]{\langle #1 \vert #2 \rangle} 
\newcommand{\id}{\mathrm{Id}}
\newcommand{\+}{\dagger}

\newcommand{\U}{\mathcal{U}}
\newcommand{\Hi}{\mathcal{H}}
\newcommand{\Iso}{\mathcal{I}}
\newcommand{\Ort}{\mathcal{O}}
\newcommand{\A}{\mathcal{A}}
\newcommand{\R}{\mathbb{R}}
\newcommand{\C}{\mathbb{C}}
\newcommand{\W}{\mathcal{W}}
\newcommand{\inj}{\mathbb{I}}
\newcommand{\proj}{\mathbb{P}}

\DeclareMathOperator{\tr}{\mathrm{tr}}
\DeclareMathOperator{\Span}{\mathrm{span}}
\def\eps{\varepsilon}

\noitemsep
\begin{document}

\author{
Fr\'ed\'eric Magniez
\thanks{CNRS--LRI, Universit\'e Paris-Sud, 91405 Orsay,
France.Partially supported by the EU 5th framework program RESQ
IST-2001-37559, and by ACI Cryptologie CR/02 02 0040 and ACI
S\'ecurit\'e Informatique 03 511 grants of the French Research
Ministry.  Part of the research was done while visiting Perimeter
Institute at Waterloo, ON, Canada. \texttt{magniez@lri.fr}}
\and Dominic Mayers
\thanks{Institute for Quantum Information, California Institute of
Technology, USA. {\tt dmayers@cs.caltech.edu}}
\and 
Michele Mosca
\thanks{University of Waterloo and Perimeter Institute, Waterloo, ON,
Canada. Partially supported by NSERC, ARDA, ORDCF, CFI and CIAR. {\tt
mmosca@iqc.uwaterloo.ca}}
\and 
Harold Ollivier
\thanks{Perimeter Institute, Waterloo, ON, Canada. Partially supported
by ACI S\'ecurit\'e Informatique, R\'eseaux Quantiques. {\tt
harold.ollivier@polytechnique.org}}
}

\title{Self-Testing of Quantum Circuits}

\maketitle

\begin{abstract}
We prove that a quantum circuit together with measurement apparatuses
and EPR sources can be fully verified without any reference to some
other trusted set of quantum devices. Our main assumption is that the
physical system we are working with consists of several identifiable
sub-systems, on which we can apply some given gates locally.

To achieve our goal we define the notions of simulation and
equivalence.  The concept of simulation refers to producing the
correct probabilities when measuring physical systems.  To enable the
efficient testing of the composition of quantum operations, we
introduce the notion of equivalence. Unlike simulation, which refers
to measured quantities (i.e., probabilities of outcomes), equivalence
relates mathematical objects like states, subspaces or gates.

Using these two concepts, we prove that if a system satisfies some
simulation conditions, then it is equivalent to the one it is purposed
to implement. In addition, with our formalism, we can show that these
statements are robust, and the degree of robustness can be made
explicit (unlike the robustness results of~\cite{dmms00}). In
particular, we also prove the robustness of the EPR
Test~\cite{my98}. Finally, we design a test for any quantum circuit
whose complexity is linear in the number of gates and qubits, and
polynomial in the required precision.
\end{abstract}

\section{Introduction}\label{intro}

We develop techniques for verifying the operations that a given set of
quantum gates perform. We consider ``self''-tests, which are tests
using the given set of gates without reference to some other trusted
and already characterized quantum devices. This notion was initially
defined for classical programs~\cite{bk95,blr93}.  Self-testing was
then extended to quantum devices~\cite{my98,dmms00} and to quantum
testers of logical properties~\cite{bfnr03,fmss03}.

The work by Mayers and Yao~\cite{my98} focuses on testing entangled
EPR states shared between two distinguishable locations, $A$ and
$B$. Apart from assuming the standard axioms of quantum mechanics, the
main assumptions they exploit are locality in the sense that the
measurements at $A$ commute with the measurements at $B$ (i.e., no
instantaneous signaling); and that one can perform independent
repetitions of the same experiments, in order to gather statistics
(i.e., the apparatuses have no memory of previous runs of the
experiments). However, they do not
assess the robustness of their results (i.e., they do not claim that
if the state satisfies the required statistics with precision $\eps$
then the state is within $\eps^{\Omega(1)}$ of an EPR
state). Robustness is nonetheless an interesting property very much
worth studying for practical reasons: first, one can never learn any
statistics with infinite precision by sampling only; second, by their
very nature, physical implementations are only approximate.

The work of Van Dam, Magniez, Mosca and Santha~\cite{dmms00} focuses
instead on testing gates. They make a number of assumptions, including
(and in addition to assuming the standard axioms of quantum mechanics)
(1) the ability to repeat the same gate in the same experiment; (2)
the absence of memory in the apparatus between different experiments;
(3) the ability to prepare and measure `0' and `1'; (4) the locality
of each of the gates (i.e., they only affect the qubits they are
suppose to act upon); and (5) the dimension of the physical
qubits (i.e., 2-level systems).  Of these assumptions, the last one is
certainly the most unrealistic one, but also the most crucial
one. Relaxing it allows for ``conspiracies'' that can spoof the test,
and it is not so clear how to work around them (See
Appendix~\ref{example} for an example given by Wim van Dam).

This paper improves upon the Mayers and Yao results~\cite{my98} by
making them robust.  It also improves upon the Van Dam, Magniez, Mosca and
Santha paper~\cite{dmms00} by removing the need for assumptions (1),
(3) and (5). Some version of assumption (2) seems necessary. We
suspect, one might be able to relax assumption (4) to some extent, but
we keep a version of it in this work.

We have sketched the assumptions of the previous work that we do not
wish to make. Let us now detail the assumptions that we do make. We
assume that, (H1) the physical system we are working with consists of
several identifiable sub-systems; (H2) two subsystems interact only if
we are applying a gate that has both those subsystems as input; (H3)
each gate will behave identically in each experiment it is used in
(i.e., each gate is some fixed completely positive superoperator); and
(H4) classical computation and control are perfect and can be trusted
(e.g., classical control has no side-channel).

Our procedure allows us to test physical implementations of unitary
gates, EPR creation gates, and one-qubit projective measurements. A
more general superoperator can be tested by viewing it as a
composition of operations of the above form. 

There is however one important restriction to the class of gates we
are able to test. The ideal gates must have real valued
coefficients. Note that we are not making any assumptions about the
physical implementation of gates, but rather on the ideal gates they
are supposed to simulate. We are merely saying that we do not have a
procedure for verifying that a physical gate is equivalent to a
complex gate. This is not for a lack of trying.  The problem is that
any complex gate of dimension $d$ can be simulated using quantum
systems of dimension $2d$, real gates and appropriate measurement
devices, in a rather standard way~\cite{RG02a}. On the positive side,
this remark means that our restriction is not a limitation.  But, this
also means that one cannot tell if a gate is complex or simulated by a
real one without external help (e.g., knowledge of the dimensionality
of quantum systems, trusted one-qubit measuring apparatuses, etc.).
More importantly, given any set of quantum gates, and a set of
experiments attempting to characterize those gates, there is a
corresponding set of real gates that would produce identical
predictions. However, these two sets of gates are not equivalent
according to the natural notion of equivalence we define (which is a
type of local unitary equivalence). That is, we believe that such
gates cannot be trusted in a cryptographic context without further
assumption. The reason is that although the real-gate simulations of
the complex gates yield identical outcome probabilities, an adversary
might be able to take advantage of the structure of real-state
simulations in order to extract information on the quantum operations
being performed.

Our first contribution (Section~\ref{concepts-section}) is to propose
a theory of self-testing by introducing appropriate notions such as
simulation and equivalence. Unlike simulation, which refers to
measured quantities (i.e., probabilities of outcomes), equivalence
relates mathematical objects like states, subspaces or
gates. Equivalence is meant to relate objects with similar observable
properties. Therefore, we have based this notion on the existence of
unitary transformations that map states and operations onto their
respective ideal version. Our notion preserves the inner product and
hence the distinguishability of quantum states, which is a crucial
tool for assessing the security of physical implementations of most
quantum cryptographic protocols.

Our second contribution (Section~\ref{theorems-section}) is a
characterization of unitary gates and circuits.  Namely, we
explain how simulation implies equivalence. The main tool for
thwarting conspiracies is the Mayers-Yao test of an EPR pair. We
will build upon the fact that one way of preparing trusted random
BB84 states is to first prepare an EPR state, transmit one half,
and independently measure the other half. We will show that this
method can be generalized and yields trusted input states to be
used in conjunction with self-testable quantum circuits.

Our last contribution (Section~\ref{test-section}) is to prove the
robustness of our characterization. In particular, we show that the
EPR test of~\cite{my98} is robust. Using the concepts of simulation
and equivalence, such proofs are not so difficult although the
robustness of the EPR test had been left open. The crucial point was
to realize that the robustness of our characterization needs only to
be stated on a rather small subspace in order for it to be of
practical interest. 

The important consequence of our study is the possibility of defining
a tester (Section~\ref{circuit-section}) %, \textbf{Circuit Test}, 
that might be used in real-life
situations. Contrary to tomography which requires trusted
measurement devices and an exponential number of statistics to be
checked, %\textbf{Circuit Test} 
our test has a complexity linear in the number
of qubits and gates involved in the circuit, and polynomial in the
required precision. We describe our tester with an example
and in a general context.
%(Section~\ref{test-section}), and 
%(Appendix~\ref{example-circuit}).

\section{Testing Concepts}\label{concepts-section}

\subsection{Notation}
In this section, we describe our theory of testing using a fixed
integer $N$ as parameter. Later in the paper, we will set $N=2$ as it
will correspond to the case of qubits.
For an introduction to quantum computing, we refer the reader to~\cite{nc00,ksv02}.

We denote by $\U(N)$ the set of unitary matrices of size $N$, $\U(H)$
the set of unitary transformations on the Hilbert space $H$, and
$\Iso(H,H')$ the set of isomorphisms between the Hilbert spaces $H$
and $H'$ (with same dimension) which preserve the inner product.  In
case of transformations over real spaces, we use the notations
$\Ort(N)$ and $\Ort(H)$ instead of $\U(N)$ and $\U(H)$.

For the Hilbert space $\Hi_2$ we denote by $\ket 0$ and
$\ket{1}$ the computational basis, and for any $\alpha
\in \R$ the state $\ket\alpha = \cos \alpha \ket 0 + \sin \alpha
\ket{1}$. In particular $\ket{\tfrac{\pi}{2}}=\ket{1}$.
We denote by $\ket{\phi^+}$ the EPR state
$\frac{1}{\sqrt 2}(\ket 0\otimes\ket 0 + \ket{1}\otimes
\ket{1})$. For $n$ finite, we denote by $\ket{\Phi^+_n}$
the state corresponding to $n$ EPR states: $\ket{\Phi^+_n} =
\frac{1}{\sqrt {2^n}} \sum_{x \in \{0,1\}^n} \ket x
\otimes \ket x$.

For linear transformations $M$ and $M'$ on $H$, and a subspace $S$ of $H$,
the notation $M=_S M'$
means that the equality holds only on $S$.  When $M$ is a linear
transformation on $A$, we extend $M$ on any tensor product $A\otimes
B$ by $M\otimes \id_B$; we sometimes still denote this as $M$ in an attempt
to simplify our notation.

\subsection{Simulation}
The concept of simulation formalizes the idea of producing the correct
probabilities when observing physical systems. Observations are based
on fixed experimental setups comprising measuring devices that gather
information about the state of the system of interest.  Likewise, the
simulation of a state by another one will be defined with respect to
projectors.  These projectors are used here in the same way
measurement devices are used in a laboratory: they act as reference
systems against which the system of interest is tested.

More precisely, we are given a family of projectors
$(P^w)_{w\in\W}$ acting on $H$, and a state $\ket\psi$ whose purpose
is to simulate the state $\ket\phi$ of the canonical Hilbert space
$\Hi_N = \C^N$.  In the following definition, and throughout the rest
of this paper, we implicitly use the labels $w$ of the projectors
$P^w$ on $H$ to label some projectors $\ketbra w$ on $\Hi_N$, that are
assumed to be given and fixed.
\begin{definition}
A quantum state $\ket{\psi}\in H$ {\em simulates} the quantum state
$\ket{\phi}\in \Hi_N$ (with respect to $\{P^w\}_{w\in\W}$),
if $\norm{ P^w \ket{\psi}}^2=\abs{\braket{w}{\phi}}^2$, for every $w\in\W$.
\end{definition}

The notion of simulation can be rephrased for a whole Hilbert space
$H$. Let $(\ket{i})_i$ be the canonical orthogonal basis of $\Hi_N$,
usually the computational basis. Assume we are given a family of
states $(\ket{\psi_i})_i$ of $H$ such that each $\ket{\psi_i}$
simulates $\ket{i}$ (with respect to fixed set of projectors
$\{P^w\}_{w\in\W}$). In such case, we say that $(\ket{\psi_i})_i$
simulates $(\ket{i})_i$ or, when there is no ambiguity, that $H$ {\em
simulates} $\Hi_N$.

With this definition of simulation for Hilbert spaces, it is possible
to extend the notion of simulation to gates. Note in the definition
below that the set of projectors used to assess that $H$ simulates
$\Hi_N$ is the same as the one used to assess that $G\ket{\psi_i}$
simulates $T\ket{i}$.
\begin{definition}
Assume that $H$ {simulates} $\Hi_N$: $(\ket{\psi_i})_i$ simulates
$(\ket{i})_i$ (with respect to $\{P^w\}_{w\in\mathcal{W}}$). A unitary
transformation $G\in\U(H)$ {\em simulates} the unitary transformation
$T\in\U(\Hi_N)$, if $G\ket{\psi_i}$ simulates $T\ket{i}$ (with respect
to $\{P^w\}_{w\in\W}$), for every $i$.
\end{definition}

\subsection{Equivalence}

One goal of testing is to ensure, using few resources, that a physical
implementation of a circuit is faithful enough so that the
probabilities for the final measurement outcomes are identical to
those that would be obtained after running the ideal
circuit. Unfortunately, the notion of simulation as defined earlier
does not compose. That is, measuring probabilities for parts of the
circuit does not guarantee that the whole will function according to
its ideal specifications. To be able to compose statements, we
introduce the notion of equivalence.

Clearly, we want a notion of equivalence that respects the inner product of
quantum states and that preserves the tensor product structure of the
different registers.
The first requirement follows from the fact that we want to be able to
conclude that equivalence implies simulation and leads to an equivalence
notion based on isometries or unitary transformations.
The second requirement is imposed in order to keep a track of local
transformations.
This is crucial in this work since a series of local tests based on EPR
pairs will be designed
in order to test a whole circuit given by a sequence of local gates.
It can be seen quite simply through the following
example that using only isometries or unitary transformations does not
satisfy this last property.

Consider two $4$-dimensional vector spaces $A$ and $B$,
and $H=A\otimes B$.
We identify in $A$ (resp. $B$) two $1$-qubit registers that we denote by
$A_1$ and $A_2$ (resp. $B_1$ and $B_2$).
Let $\ket{\psi}=\ket{\phi^+}_{A_1B_1}\otimes\ket{\phi^+}_{A_2B_2}$.
If the measurements on $A$ (resp. $B$) only measure the $A_1$-part of $A$
(resp. the $B_1$-part of $B$),
we would like to say that $\ket{\psi}$ is equivalent to $\ket{\phi^+}$
on the subspace $S=\{\ket{\varphi}_{A_1B_1}\otimes\ket{\phi^+}_{A_2B_2} :
\ket{\varphi}_{A_1B_1}\in A_1\otimes B_1\}$,
since the $(A_2\otimes B_2)$-part of the system is not used.
Even if there exists an isometry $U\in\Iso(S,\Hi_4)$ such that
$U\ket{\psi}=\ket{\phi^+}$ (and $P^{a,b}=_S U^\+\ketbra{a,b} U$)
this isometry cannot be decomposed with respect to the tensor decomposition of $H$.
However this is fundamental for our purposes.
This justifies a more elaborated notion of equivalence where we introduce
a logical counterpart to any Hilbert space.

The equivalence notion we now introduce is based on the work of Mayers and Yao~\cite{my98}.
It is a mathematical notion based 
on the possibility of transferring states which lie within a given subspace
of $H$ into a logical system $H_c$ prepared in a fiducial state via
a joint unitary transformation.

For a Hilbert space $H$, that will describe the state of our physical
system,
we set a {\em logical} space $H_c=\Hi_N$ and define $\bar H = H_c \otimes H$.
We consider in $H_c$
the usual canonical basis $(\ket{i})_{0\leq i < N}$, so that we have a
canonical mapping between $H_c$ and $\Hi_N$, between $\U(H_c)$ and
$\U(N)$, and between $\Ort(H_c)$ and $\Ort(N)$.
Note that it is more convenient to set this logical system outside the physical system
(instead of as a subpart of it) since initially we do not know which part of
the physical system is used for the computation. Identifying some subsystem
of $H$ as the logical space seems more unnatural than just adding this
additional logical qubit.

The state $\ket \psi\in H$ is embedded in $\bar H$ using the isometry:
$\inj_{ H}: \ket \psi \mapsto \ket{0} \otimes \ket\psi$.
The reverse operation is obtained by applying:
$\proj_{ H}: \ket \psi \mapsto \tr_{\Hi_N}
((\ketbra{0} \otimes \id_H) \ket \psi)$.
It can be checked that $\proj_{ H}  \inj_{ H} = \id_H$. 
The operators $\proj_H$ and $\inj_H$ allow to 
identify $H$ with the subspace $\ket 0 \otimes
H$ of $\bar H$. Similarly, any linear map $M$ on $H$ is extended to
the linear map $\ketbra 0 \otimes M$ on $\bar H$. Thus, we will omit
$\proj_H$ and $\inj_H$ when there is no ambiguity.

First, we define the equivalence between a subspace of $H$ and the
logical system $H_c$ with respect to a set of projectors. As for the
notion of simulation, these projectors act as reference systems.
\begin{definition}\label{def_eq}
Let $U\in\U(\bar H)$. A subspace $S$ of $H$ is {\em $U$-equivalent} to
$H_c$ (with respect to $(P^w)_{w\in\mathcal{W}}$), if
for every $w\in\mathcal{W}$,
$P^w =_{S} \proj_{ H}  U^\+  (\ketbra{w} \otimes \id_H)
 U \inj_{ H}$.
\end{definition}
The above definition is equivalent to the commutative diagram:
$$
\begin{array}{rcl}
S &\xrightarrow{P^w}& S \\
U \inj_{ H}\downarrow & & \uparrow \proj_{ H} U^\+\\
\bar H &\xrightarrow{\ketbra{w}\otimes\id_H} & \bar H 
\end{array}
.$$

Intuitively, the unitary transformation $U$ ensures that the
correspondence between the physical system $H$ and the 
logical system $H_c$ is well defined on $S$.
Using this correspondence, we can now define the notion of $U$-equivalence 
for states and gates. 
\begin{definition}
Let $S$ be a subspace of $H$. A state $\ket{\psi}\in S$ is {\em
$U$-equivalent} to a state $\ket{\phi}\in H_c$ on $S$ (with respect to
$(P^w)_{w\in\mathcal{W}}$), if
\vspace*{-7pt}\begin{enumerate}
\item $S$ is $U$-equivalent to $H_c$,
\item $\ket{\psi} = U^\+ (\ket{\phi}\otimes\ket{\chi})$, for some
$\ket{\chi} \in H$.
\end{enumerate}\vspace*{-7pt}
\end{definition}

\begin{definition}
Let $S$ be a subspace of $H$.
A unitary transformation $G\in\U(H)$
is {\em $(U,V)$-equivalent} 
to a unitary transformation $T\in\U(H_c)$ on $S$
(with respect to $(P^w)_{w\in\mathcal{W}}$), if
\vspace*{-7pt}\begin{enumerate}
\item $S$ is $U$-equivalent to $H_c$,
\item $S'=G(S)$ is $V$-equivalent to $H_c$,
\item $G =_{S} V^\+ (T\otimes W) U$, for some $W \in \U(H)$.
\end{enumerate}\vspace*{-7pt}
\end{definition}

This equivalence can be summarized by the following commutative
diagram:
$$
\begin{array}{rcccccl}
S &\xrightarrow{P^w}& S & \xrightarrow{G}& S'&\xrightarrow{P^w}& S'\\
U \inj_{ H}\downarrow & &  \proj_{ H} U^\+\uparrow
\downarrow U \inj_{ H}& &  \proj_{ H} V^\+ \uparrow
\downarrow V \inj_{ H}& & \uparrow \proj_{ H} V^\+\\
\bar H &\xrightarrow{\ketbra{w}\otimes\id_H} & \bar H 
&\xrightarrow{T\otimes W} & \bar H
&\xrightarrow{\ketbra{w}\otimes\id_H} & \bar H 
\end{array}.
$$

When $H$ is explicitly decomposed into a tensor product, $H =
\bigotimes_{i=1}^n H^i$, and $P^w=\bigotimes_{i=1}^n P_{H^i}^{w^i}$,
where $w=(w^1,w^2,\ldots,w^n) \in \mathcal W^1 \times \mathcal W^2
\ldots \mathcal W^n $, we will often use the notion of equivalence for
unitary matrices $U$ that can be tensor product decomposed as
$U=\bigotimes_i U^i$, for some $U^i\in \U(\bar H^i)$.  When we do not
want to specify the decomposition of $U$, we will use the notion of
{\em tensor equivalence}.  Notice that for the state and
transformation tensor equivalence, $\ket{\chi}$ and $W$ are not
required to be tensor product decomposable.
This is because we want
%Thus, 
%equivalence
encompass situations where the physical implementation $G$ of the
gate creates or destroys entanglement in the hidden degrees of freedom
of the quantum register.

Finally, note that the tensor equivalence on $H$ implies the
equivalence for each factor $H_i$ of the tensor decomposition of $H$,
if for each factor $H_i$ one can sum up some projections
$P_{H^i}^{w^i}$ to the identity. This will be the case in the rest of
the paper.
\begin{proposition}
Let $H=\bigotimes_{i=1}^n H^i$.  Let $S$ be a subspace of $H$ which is
$(\bigotimes_i U_i)$-equivalent to $H_c=\bigotimes_i {H^i_c}$ with
respect to $(P^w)_w$.  Assume that for every $i$, a subset of the
projectors of $(P_{H^i}^{w^i})_{w^i \in \mathcal W^i}$ sums to the
identity on $H^i$.  Then $S$ is $U_i$-equivalent to $H_c$ with respect
to $(P_{H^i}^{w^i})_{w^i \in \mathcal W^i}$, for every $i$.  Moreover
if $S=\bigotimes_i S^i$, where $S^i$ is a subspace of $H^i$, then
$S^i$ is $U_i$-equivalent to $H^i_c$ with respect to
$(P_{H^i}^{w^i})_{w^i \in \mathcal W^i}$, for every $i$.
\end{proposition}

From now on, we set $N=2$ when we do not explicitly state
otherwise. When we omit the parameters $U$ or $(U,V)$ from the
equivalence notation, we mean that there exists such unitary
transformations for which the $U$-equivalence or the
$(U,V)$-equivalence holds.

\subsection{EPR Test}
In this section, we summarize Mayers and Yao's results~\cite{my98} in
the framework of quantum testing we have just introduced. Their main
result {\cite[Thm. 1]{my03}} will be stated in an extended form that
is most convenient for testing several registers successively.

{From} now and until the end of the paper, let
$\A_0=\{0,\tfrac{\pi}{8},\tfrac{\pi}{4}\}$, $\A_1=\{ a+\tfrac{\pi}{2}:
a\in \A_0\}$, and $\A=\A_0\cup \A_1$. We fix in this section
$(P_A^a,P_A^{a+\pi/2})_{a\in \A_0}$ and $(P_B^b,P_B^{b+\pi/2})_{b\in
\A_0}$ orthogonal measurements respectively on two Hilbert spaces $A$
and $B$. Namely, we assume that $P_A^a+P_A^{a+\pi/2}=\id_A$ and
$P_B^a+P_B^{a+\pi/2}=\id_B$, for every $a\in\A_0$.

\begin{theorem}\label{MY-theorem}
Let $H = A\otimes B\otimes C$, and $\ket{\psi}\in H$ that simulates
$\ket{\phi^+}$ with respect to $(P_A^a\otimes P_B^b\otimes \id_C)_{a,b
\in\A}$.  Then there exist two unitary transformations $U_{\bar
A}\in\U(\bar A)$ and $U_{\bar B}\in\U(\bar B)$ such that $\ket{\psi}$
is $(U_{\bar A}\otimes U_{\bar B})$-equivalent to $\ket{\phi^+}$ on $S
= \Span\{P_A^a \otimes P_B^b\otimes\id_C\ket\psi: a,b \in \A\}$.
Moreover the dimension of $S$ is 4.
\end{theorem}
Note that the theorem can be extended from $S$ to the supports of
$\ket{\psi}$ on the $A$-side and on the $B$-side using
\cite[Prop. 4]{my03}. Since we will only need the result on $S$, and
because the robustness the EPR test is easier to state in such case,
we will only state our results for $S$, even though all of them can be
extended to the tensor product of the respective supports (for the
exact case).

{From} Theorem~\ref{MY-theorem} it is easy to derive by induction over
$n$ our main tool for testing $n$-qubit registers. Let $A =
\bigotimes_{i=1}^n A^i$ and $B = \bigotimes_{i=1}^n B^i$, we now fix
$(P_{A^i}^{a^i},P_{A^i}^{a^i+\pi/2})_{a^i \in \A_0}$ and
$(P_{B^i}^{b^i},P_{B^i}^{b^i+\pi/2})_{b^i \in \A_0}$ to be orthogonal
measurements on $A^i$ and $B^i$ respectively for every $i$. We denote
$P_A^{a} = \bigotimes_{i=1}^n P_{A^i}^{a^i}$, with $a = (a^i)_{i=1}^n$
and $P_B^{b} = \bigotimes_{i=1}^n P_{B^i}^{b^i}$ with $b =
(b^i)_{i=1}^n$. Note that in the following corollary, the tensor
equivalence is with respect to the tensor decomposition $A\otimes B$,
but also with respect to the tensor decompositions $A =
\bigotimes_{i=1}^n A^i$ and $B = \bigotimes_{i=1}^n B^i$.
\begin{corollary}\label{MY-cor1}
Let $H = A\otimes B\otimes C$, and $\ket{\Psi}\in H$ that simulates
$\ket{\phi^+}$ with respect to $(P_{A^i}^{a^i}\otimes
P_{B^i}^{b^i}\otimes\id_C)_{a^i,b^i \in\A}$ for every
$i=1,2,\ldots,n$.  Then there exist two unitary transformations
$U_{\bar A}\in \bigotimes_i \U(\bar A^i)$ and $U_{\bar B}\in
\bigotimes_i \U(\bar B^i)$ such that $\ket{\Psi}$ is $(U_{\bar
A}\otimes U_{\bar B})$-equivalent to $\ket{\Phi^+_n}$ on
$S = \Span\{P_A^a \otimes P_B^b\ket\psi: a,b \in \A^n\}$.
Moreover the dimension of $S$ is $4^n$.
\end{corollary}

Therefore, when measurements are acting on different factors of the
tensor product decompositions of $A$ and $B$, testing a $2n$-qubit EPR
state can be done by testing the $n$ EPR pairs that are present in
it. That is, by checking the probabilities of $O(n)$ outcomes, whereas
there are $2^{O(n)}$ possible joint measurement outcomes.

\section{Simulation implies Equivalence}\label{theorems-section}

In this section we relate simulation and equivalence. While it is
clear that equivalence implies simulation, we show below that under
certain assumptions, simulation implies equivalence. To ease the
presentation of our results, we start by describing how $1$-qubit real
gates, namely transformations in $\Ort(2)$, can be tested. As a second
step, we show how to test $n$-qubit real gates.

\subsection{One-qubit Gate Testing}

As a first attempt, we show how to test that a gate is acting
as the identity. 
\begin{proposition}
Let $H = A \otimes B$ and $G \in \U(A)$. Let $\ket\psi \in H$ be such
that $\ket \psi$ and $G\ket\psi$ simulate $\ket{\phi^+}$. Then,
$G\otimes\id_B$ is tensor equivalent to $\id_{A_c}\otimes\id_{B_c}$ on
$S = \Span\{P_A^a \otimes P_B^b\ket\psi: a,b \in \A\}$.
\end{proposition}

\begin{proof}
We show below that $G$ is $(U_{\bar A}\otimes U_{\bar B}, U_{\bar
A}G^\+ \otimes U_{\bar B})$-equivalent to $\id_{A_c}\otimes \id_{B_c}$
on $S$.

First note that Lemmas~\ref{MY-lemma1} and~\ref{MY-lemma2} applied to
$\ket\psi$ gives $U_{\bar A}$ and $U_{\bar B}$ such that $S$ is
$(U_{\bar A}\otimes U_{\bar B})$-equivalent to $A_c\otimes B_c$ and
$U_{\bar A}\otimes U_{\bar B} \ket \psi = \ket{\phi^+}\otimes
\ket\chi$ for some $\ket\chi$ in $A\otimes B$. We can derive
that $(U_{\bar A} G^\+ \otimes U_{\bar B}) G \ket\psi = \ket{\phi^+}
\otimes \ket\chi$.

Hence, it only remains to show that $G(S)$ is $(U_{\bar A} G^\+
\otimes U_{\bar B})$-equivalent to $A_c\otimes B_c$.  Let $a,b,a',b'
\in \A$, then the following equalities hold:
\begin{align*}
(P_A^a\otimes P_B^b) (G\otimes\id_B)(P_A^{a'}\otimes P_B^{b'})
\ket\psi 
& = (\id_A\otimes P_B^b P_B^{b'} P_B^{a'}) (P_A^{a'} G\otimes\id_B)
\ket\psi \\
& = G \otimes P_B^b P_B^{b'} P_B^{a'} P_B^a \ket \psi \\
& = (G \otimes \id_B)(P_A^a\otimes P_B^b)(P_A^{a'}\otimes
P_B^{b'})\ket\psi,
\end{align*}
where we applied Proposition~\ref{MY-prop1} (see Appendix~\ref{my3}) to $\ket\psi$ on the first
and the last line, and to $G\ket\psi$ on the second line. In other
words, this states that $(P_A^a \otimes P_B^b) (G\otimes\id_B) =_{S}
(G\otimes\id_B)(P_A^a \otimes P_B^b)$.  Using $U_A\otimes U_B$ to
replace $P_A^a\otimes P_B^b$ over $S$, we get
$P_A^a \otimes P_B^b =_{G(S)} (G U_{\bar A}^\+ \otimes U_{\bar B}^\+)
(\ketbra a \otimes \ketbra b) (U_{\bar A}G^\+ \otimes U_{\bar B})$,
which is the required equivalence between $G(S)$ and $A_c\otimes B_c$.
\end{proof}
Stating the above result allows us to exhibit simple characteristics
of the general method used for proving that gates can be
self-tested. First, any gate testing requires two EPR tests. These are
used to ensure that the input and output states together with the
measurements act properly before and after the gate. These are
``conspiracy'' tests. Second, the fundamental properties of EPR
states---namely that a given measurement can be performed on either
the $A$-side or the $B$-side without changing the collapsed state---is
used in order to show that on the input state $\ket\psi$, the gate $G$
and the measurements commute. Together with the replacement of the
projectors $P_A^a$ and $P_B^b$, that come from the physical measurements,
by their ideal versions $\ketbra a$ and $\ketbra b$ on $A_c$ and $B_c$, this allows to perform
the tomography of the gate $G$.

We can now state the general result concerning any $1$-qubit real
gate.
\begin{theorem}\label{1qubit-theorem}
Let $T \in \Ort(2)$.  Let $H = A \otimes B$, $G_A \in \U(A)$, and
$G_B\in\U(B)$.  Let $\ket\psi \in H$ be such that $\ket \psi$ and $G_A
G_B\ket\psi$ simulate $\ket{\phi^+}$, and such that $G_A\ket\psi$
simulates $(T\otimes \id_2)\ket{\phi^+}$.  Then, $G_A$ is tensor
equivalent to $T$ on $S = \Span\{P_A^a\otimes P_B^b\ket\psi : a,b \in
\A\}$.
\end{theorem}

\begin{proof}
The proof proceeds in two steps. First, it is shown that $S$ and
$G_A(S)$ are respectively $(U_{\bar A}\otimes U_{\bar B})$- and
$(V_{\bar A}\otimes U_{\bar B})$-equivalent to $A_c\otimes
B_c$. Second, it is shown that there exists $W \in \U(A)$ such that
$G_A\otimes\id_B =_S (V_{\bar A}^\+ \otimes U_{\bar B}^\+) (T\otimes W
\otimes \id_{\bar B}) (U_{\bar A} \otimes U_{\bar B})$.

Lemmas~\ref{MY-lemma1} and \ref{MY-lemma2} applied to $\ket\psi$ and
$G_AG_B\ket\psi$ give $U_{\bar A}, V_{\bar A} \in \U(\bar A)$ and
$U_{\bar B}, V_{\bar B} \in \U(\bar B)$ such that $S$ and $(G_A\otimes
G_B) (S)$ are respectively $(U_{\bar A}\otimes U_{\bar B})$- and
$(V_{\bar A}\otimes V_{\bar B})$-equivalent to $A_c\otimes B_c$.  This
implies that $(G_A\otimes\id_B)(S)$ is $(V_{\bar A}\otimes U_{\bar
B})$-equivalent to $A_c\otimes B_c$. That is, we have the required
tensor equivalences for $S$ and $G_A(S)$. If we define $\ket
\chi_{AB}$ as $U_{\bar A}\otimes U_{\bar B} \ket\psi =
\ket{\phi^+}_{A_cB_c}\otimes \ket\chi_{AB}$, we then have
$S=U_A^\+\otimes U_B^\+ (A_c\otimes B_c\otimes {\ket{\chi}_{AB}})$.

The simulation of $T\ket{\phi^+}$ by $G_A\ket\psi$ can be rewritten
within the density matrix formalism as: $\tr \left( (P_A^a \otimes
P_B^b) G_A\ketbra{\psi}G_A^\+\right) = \tr \left( (\ketbra{a}\otimes
\ketbra{b}) (T\otimes \id_2) \ketbra{\phi^+} (T^\+\otimes
\id_2)\right)$.  Using the commutativity of the trace operator and
$(\id_2 \otimes \ketbra b)\ketbra{\phi^+} = \tfrac{1}{2} \ketbra{b}
\otimes \ketbra{b}$, we get $\tr\left( (G_A^\+ P_A^a G_A \otimes P_B^b
) \ketbra\psi \right) = \frac{1}{2} \tr\left(T^\+ \ketbra a T \ketbra
b \right)$.

Define the positive semi-definite operator $R^a_{\bar A\bar B} =
(U_{\bar A} \otimes U_{\bar B}) G_A^\+ P_A^a G_A (U_{\bar A}^\+\otimes
U_{\bar B}^\+)$. Since $\ket\psi$ is tensor equivalent to
$\ket{\phi^+}$, we have: $\tr\left(R^a_{\bar A\bar B} (\ketbra b_{A_c}
\otimes \ketbra b_{B_c} \otimes \ketbra\chi_{AB}) \right) =
\tr\left(T^\+ \ketbra a T \ketbra b \right)$.

This can easily yield the equations required to apply
Lemma~\ref{lem:tomo_N} for performing the tomography of $R^a_{\bar
A\bar B}$. For instance, observe that the operators $U_{\bar B}$ and
$U_{\bar B}^\+$ can be removed from the definition of $R^a_{\bar A\bar
B}$ without modifying it.  Therefore the previous equation can be
extended for all $b,b' \in \A$ to
$$\tr\left(R^a_{\bar A\bar B} (\ketbra b_{A_c} \otimes
\ketbra{b'}_{B_c} \otimes \ketbra\chi_{AB}) \right) = \tr\left(T^\+
\ketbra a T \right),$$ since the value of the left hand side does not
depend on $b'$.

Now Lemma~\ref{lem:tomo_N} can be applied on $A_c$ to the operators
${}_{AB}\bra{\chi}{}_{B_c}\bra{b'}R^a_{\bar A \bar
B}\ket{b'}_{B_c}\ket{\chi}_{AB}$ and $T^\+\ketbra{a}T$ with $n=1$ and
$\eps=0$.  The conclusion is that
${}_{AB}\bra{\chi}{}_{B_c}\bra{b'}R^a_{\bar A \bar
B}\ket{b'}_{B_c}\ket{\chi}_{AB}=(T^\+\ketbra{a} T)$, for every
$b'\in\A$.  Since $R^a_{\bar A \bar B}$ is a semi-definite operator,
the above conclusion can be rewritten as
\begin{equation}
R^a_{\bar A \bar B} =_{A_c\otimes B_c\otimes {\ket{\chi}_{AB}}} (T^\+
 \ketbra a T)\otimes \id_{A \otimes {\bar B}}.\label{eq:afterlemma}
\end{equation}

The tensor-equivalence of $G_A(S)$ with $A_c\otimes B_c$ also gives 
\begin{equation*}
P_A^a  =_{G_A(S)} (V_{\bar A}^\+ \otimes U_{\bar
B}^\+)(\ketbra a \otimes \id_{A \otimes {\bar B}})(V_{\bar A}
\otimes U_{\bar B}).
\end{equation*}
Since $S=U_A^\+\otimes U_B^\+ (A_c\otimes B_c\otimes {\ket{\chi}})$,
this can be used to replace $P_A^a$ inside Equation~\eqref{eq:afterlemma}.
We obtain
\begin{multline*}
(\ketbra a \otimes \id_{A \otimes {\bar B}}) (V_{\bar A} \otimes
U_{\bar B}) G_A(U_{\bar A}^\+ \otimes U_{\bar B}^\+) (T^\+\otimes
\id_{A \otimes {\bar B}}) \\ 
=_{A_c\otimes B_c\otimes {\ket{\chi}}} (V_{\bar A} \otimes
U_{\bar B}) G_A (U_{\bar A}^\+ \otimes U_{\bar B}^\+)(T^\+\otimes
\id_{A \otimes {\bar B}}) (\ketbra a \otimes \id_{A \otimes {\bar B}}).
\end{multline*}
Then, we conclude using Lemma~\ref{lem:commut} with $\eps=0$, that
 there exists $W \in \U(A)$ such that
\begin{equation*}
G_A =_S (V_{\bar A}^\+ \otimes U_{\bar B}^\+)(T \otimes W \otimes
\id_{\bar B}) (U_{\bar A} \otimes U_{\bar B}).
\end{equation*}
\end{proof}

\subsection{Many-qubit Gate Testing}

We now consider $n$-qubit real gates. We present our main result for
testing gates using a slightly different formulation than in
Theorem~\ref{1qubit-theorem}. The reason for this change is that it
makes the proof of the composition theorem
(Theorem~\ref{circuit-theorem}) used for self-testing circuits
straightforward. We have also added an extra Hilbert space $C$ in the
tensor product decomposition of $H$. The proof is omitted since it is
identical to the second step of the proof of
Theorem~\ref{1qubit-theorem}, where $a$, $b$ and $b'$ are now in
$\A^n$.

\begin{theorem}\label{nqubit-theorem}
Let $T \in \Ort(2^n)$. Let $H = A \otimes B\otimes C$, where $A =
\bigotimes_i A^i$ and $B=\bigotimes_i B^i$. Let $G_A \in \U(A) $ and
$G_B \in \U(B)$.  Let $\ket \Psi \in H$ and $U_{\bar A}, V_{\bar A}
\in\bigotimes_i\U(\bar A_i)$ and $U_{\bar B}, V_{\bar B}
\in\bigotimes_i\U(\bar B_i)$ be such that:
\begin{enumerate}
\item $\ket\Psi$ is $(U_{\bar A}\otimes U_{\bar B})$-equivalent to
$\ket{\Phi^+_n}$ on $S$ with respect to $(P_A^{a}\otimes P_B^{b})_{a, b
\in \A^n}$,
\item $G_AG_B\ket\Psi$ is $(V_{\bar A}\otimes V_{\bar B})$-equivalent
to $\ket{\Phi^+_n}$ on $(G_A \otimes G_B)(S)$ with respect to
$(P_A^{a}\otimes P_B^{b})_{a, b \in \A^n}$,
\item $G_A\ket\Psi$ simulates $(T\otimes\id_{2^n}) \ket{\Phi^+_n}$ with
respect to $(P_A^{a}\otimes P_B^{b}\otimes \id_C)_{a, b \in \A^n}$,
\end{enumerate}\vspace*{-7pt}
where $S = \Span\{P_A^a \otimes P_B^b\ket\psi: a,b \in \A^n\}$.
Then $G_A$ is $(U_{\bar A}\otimes U_{\bar B}, V_{\bar
A}\otimes U_{\bar B})$-equivalent to $T$ on $S$.
\end{theorem}
Using Corollary~\ref{MY-cor1}, one can observe that this formulation
is not weaker than the one of Theorem~\ref{1qubit-theorem}.

\subsection{Circuit Testing}
Now we state our main theorem and its corollary which relates the
simulation of states to the equivalence of gates, and therefore to the
simulation of gates.  We omit their proof due to the lack of space and
because they are derived easily from Corollary~\ref{MY-cor1} and
Theorem~\ref{nqubit-theorem}.

Assume that some Hilbert space $H$ has a tensor product decomposition
$H=\bigotimes_{i=1}^n A^i\bigotimes B^i$. For any subset
$I\subseteq\{1,2,\ldots,n\}$, let $H^I$ denote the Hilbert space
$\bigotimes_{i\in I} A^i \bigotimes_{i\in I}B^i$, and
$\ket{\Phi^{+}}_{I}$ the corresponding EPR state
$\ket{\Phi^+_{|I|}}$ over $\bigotimes_{i\in I}A_c^i\bigotimes_{i\in
I}B_c^i$.

\begin{theorem}\label{circuit-theorem}
Let $H = A \otimes B$, where $A = \bigotimes_i A^i$ and
$B=\bigotimes_i B^i$.  Let $I^1,I^2,\ldots, I^t\subseteq \{1,2,\ldots,
n\}$ be $t$ subsets.  Let $G_A^j \in \U(A^{I^j})$, $G_B^j \in
\U(B^{I^j})$ and $T^j \in\Ort(A_c^{I^j})$.  Let $\ket \Psi \in
A\otimes B$.  Define inductively $\ket{\Psi'^j} =(G_A^j \otimes
\id_B)\ket{\Psi^{j-1}}$ and $\ket{\Psi^j} = (G_A^j \otimes G_B^j
)\ket{\Psi^{j-1}}$, where $\ket{\Psi^0}=\ket{\Psi'^0}=\ket{\Psi}$.
Assume the following.
\vspace*{-7pt}\begin{enumerate}
\item $\ket{\Psi}$ simulates $\ket{\phi^+}$ with respect to
$(P_{A^i}^{a^i}\otimes P_{B^i}^{b^i})_{a^i,b^i\in \A}$, for every
$i=1,2,\ldots,n$.
\item For every $j=1,\ldots,t$: $\ket{\Psi^j}$ simulates
$\ket{\phi^+}$ with respect to $(P_{A^i}^{a^i}\otimes
P_{B^i}^{b^i})_{a^i,b^i\in \A}$, for every $i\in I^j$.
\item For every $j=1,\ldots,t$: $\ket{\Psi'^j}$ simulates $T^j
\ket{\Phi^{+}}_{I^j}$ with respect to $(P_{A^{I^j}}^{a}\otimes
P_{B^{I^j}}^{b})_{a, b \in \A^{I^j}}$.
\end{enumerate}\vspace*{-7pt}
Then $G_A^t G_A^{t-1}\cdots G_A^1$ is tensor equivalent to
$T^t T^{t-1}\cdots T^1$ on 
$S=\Span(P_A^a\otimes P_B^b \ket{\Psi}: a,b\in\A^n)$.
\end{theorem}

\begin{corollary}\label{circuit-input-corollary}
Let $\ket{\Psi}\in H$ that satisfies the hypothesis of
Theorem~\ref{circuit-theorem} for some decomposition of $G_A\in\U(A)$
and $T\in\U(A_c)$ into $t$ gates acting only on a constant number of
qubits.  Then, for every $x\in\{0,1\}^n$, the state
${\sqrt{2^n}}\tr_B(P_B^x\ket{\Psi})$ simulates $\ket{x}_{A_c}$ with
respect to $(P_A^w)_{w\in\A^n}$.  Moreover $G_A$ simulates $T$ with
respect to the above identification, and the number of statistics to
be checked is in $O(t)$.
\end{corollary}

\section{Robustness of Simulation}\label{test-section}
\subsection{Norm and Notation}
We consider the $\ell_2$ norm $\norm{\cdot}$ for states, and the
corresponding operator $\norm{\cdot}$ norm for linear transformations.
These norms are stable by tensor product composition in the following
sense: $\norm{ u \otimes v}=\norm{u}\times \norm{v}$, if $u$ and $v$
denote either vectors or linear transformations.

We note $\ket{\psi}=^\eps \ket{\psi'}$ when two vectors
$\ket{\psi},\ket{\psi'}$ are such that
$\norm{\ket{\psi}-\ket{\psi'}}\leq\eps$.  We extend the
$\ell_2$-operator norm for restrictions of linear transformations on
$H$.  Namely if $M$ is a linear transformation on $H$, and $S$ is a
subspace of $H$ we define by $\norm{M}_S=\sup (\norm{M\ket{\psi}}:
\ket{\psi}\in S \text{ and } \norm{\ket{\psi}}=1)$. Similarly to
states, we will write $M=^\eps_S N$ when $\norm{M-N}_S\leq \eps$.

We introduce the notion of {\em $\eps$-simulation} by extending the
notion of simulation where statistics equalities are only
approximately valid up to some additive term $\leq \eps$. The notions
of equivalence can be similarly extended to {\em $\eps$-equivalence},
by replacing each equality $=_S$ by $=^\eps_S$.

We will not detail the multiplicative constants that will occur in the
upper bound on our additive error terms, but we will use instead the
notation $O(f(\eps))$ that denotes the existence of a universal
constant $c$ for which the upper bound $c\times f(\eps)$ is valid. We
will use the notation $\Omega(f(\eps))$ in a similar way.

\subsection{Robustness}
Until now, our interest has been focused on the possibility of
self-testing a quantum circuit when outcome probabilities are known
with perfect accuracy. To be of practical interest, our results must
be extended to the situation of finite accuracy. We show below that it
is possible and that all the relevant results for testing are indeed
robust in the following way: if the statistics are close to the ideal
ones, then the states, the measurements and the gates are also close to
ones that are equivalent to the ideal ones.  This notion of robustness
follows the ones of Rubinfeld and Sudan~\cite{rs96,rub99} for
classical computing and of~\cite{dmms00} for quantum computing.

One can extend quite easily Theorem~\ref{MY-theorem} on the vector
space $S=\Span(P_A^a P_B^b \ket\psi : a,b\in\A)$, which is enough for
our purposes.  Note that a robust version of Theorem~\ref{MY-theorem}
that would be valid on the tensor product of the supports of
$\ket\psi$ on the $A$-side and on the the $B$-side is much more
difficult to state as well as inefficient in its robustness parameter
$\eps$. This is because its conclusion might depend on the dimensions
of $A$ and $B$.

\begin{theorem}\label{a-MY-theorem}
Let $H = A\otimes B\otimes C$, and $\ket{\psi}\in H$ that
$\eps$-simulates $\ket{\phi^+}$ with respect to $(P_A^a\otimes
P_B^b\otimes \id_C)_{a,b \in\A}$.  Then there exist two unitary
transformations $U_{\bar A}\in\U(\bar A)$ and $U_{\bar B}\in\U(\bar
B)$ such that $\ket{\psi}$ is $(O({\eps}^{1/4}),(U_{\bar A}\otimes
U_{\bar B}))$-equivalent to $\ket{\phi^+}$ on $S$.
\end{theorem}

The proof can be found in Appendix~\ref{my}. This result can be
generalized to the case of a source producing a state $\ket\Psi$ that
simulates $n$ EPR pairs. In such case equivalence holds within
$O(4^n\eps)$. 
\begin{corollary}\label{a-MY-n}
Let $H = A \otimes B \otimes C$, where $A = \bigotimes_i A^i$ and
$B=\bigotimes_i B^i$. Let $\ket{\Psi}\in H$ be a state that
$\eps$-simulates $\ket{\phi^+}$ with respect to $(P_{A^i}^{a^i}\otimes
P_{B^i}^{b^i})_{a^i,b^i\in \A}$, for every $i=1,2,\ldots,n$. Then,
$\ket\Psi$ is $O(4^n\eps^{1/4})$-equivalent to $\ket{\Phi^+_n}$.
\end{corollary}
Another corollary that we will use in the context of circuit testing
concerns the case of $n$ sources of EPR pairs that are tested
simultaneously. This is qualitatively different from the previous
situation as the state $\ket\Psi$ that is tested is assumed to be
separable across the tensor product decomposition of $H$ into $H^i =
A^i \otimes B^i$.
\begin{corollary}\label{a-MY-n-sep}
Let $H = A \otimes B \otimes C$, where $A = \bigotimes_i A^i$ and
$B=\bigotimes_i B^i$. Let $\ket{\Psi}\in H$ be a separable state
across the tensor product decomposition of $H$ into $A_i \otimes B_i$,
and such that it $\eps$-simulates $\ket{\phi^+}$ with respect to
$(P_{A^i}^{a^i}\otimes P_{B^i}^{b^i})_{a^i,b^i\in \A}$, for every
$i=1,2,\ldots,n$. Then, $\ket\Psi$ is $O(n\eps^{1/4})$-equivalent to
$\ket{\Phi^+_n}$.
\end{corollary}
The proof of these two corollaries can be found in Appendix~\ref{my}.
Now we concentrate on the robustness of Theorem~\ref{nqubit-theorem}
which is proven in Appendix~\ref{a-tomography}. Note that the
exponential dependency in the number $n$ of qubits it not a
constraint, since we will use this theorem for constant $n$ only
(i.e., we assume an upper bound on the number of qubits affected by a
gate, say $n \leq 3$).
\begin{theorem}\label{a-nqubit-theorem}
Let $T \in \Ort(2^n)$. Let $H = A \otimes B\otimes C$, where $A =
\bigotimes_i A^i$ and $B=\bigotimes_i B^i$. Let $G_A \in \U(A) $ and
$G_B \in \U(B)$.  Let $\ket \Psi \in H$ and $U_{\bar A}, V_{\bar A}
\in\bigotimes_i\U(\bar A_i)$ and $U_{\bar B}, V_{\bar B}
\in\bigotimes_i\U(\bar B_i)$  be such that:
\vspace*{-7pt}\begin{enumerate}
\item $\ket\Psi$ is $(\eps,(U_{\bar A}\otimes U_{\bar B}))$-equivalent
to $\ket{\Phi^+_n}$ on $S$ with respect to $(P_A^{a}\otimes
P_B^{b})_{a, b \in \A^n}$,
\item $G_A\otimes G_B\ket\Psi$ is $(\eps,(V_{\bar A}\otimes V_{\bar
B}))$-equivalent to $\ket{\Phi^+_n}$ on $(G_A \otimes G_B)(S)$ with
respect to $(P_A^{a}\otimes P_B^{b})_{a, b \in \A^n}$,
\item $G_A\ket\Psi$  $\eps$-simulates $(T\otimes\id_{2^n})
\ket{\Phi^+_n}$ with respect to $(P_A^{a}\otimes P_B^{b}\otimes
\id_C)_{a, b \in \A^n}$.
\end{enumerate}\vspace*{-7pt}
Then $G_A \otimes \id_B$ is $(2^{O(n)}\sqrt{\eps},(U_{\bar A}\otimes
U_{\bar B}, V_{\bar A}\otimes U_{\bar B}))$-equivalent to $T\otimes
\id_{\bar B_c}$ on $S$.
\end{theorem}

\section{Testing a Circuit on a Specific Input}\label{circuit-section}
\subsection{Construction} 
The assumptions we have made so far for gate testing are allowing very
broad and generic conspiracies. For instance, the behavior of a gate
can depend on previously applied gates in the circuit. Hence, it is
impossible to have a fixed finite set of tests for characterizing the
individual gates and then trust that the composition of these gates in
a circuit will correctly simulate the ideal circuit. In other words,
any circuit used for computation must be part of some tests.

Surprisingly, it is much easier to test the simulation of a circuit on
the subspace $S$ than on a particular input. In fact, using EPR pairs
allows for the simultaneous testing of all possible inputs, while
making the selection of a particular one difficult. The obvious
choice would be to post-select the outcome of the $B$-side
measurements of the EPR pairs. Unfortunately, the selected input state
would then be prepared with exponentially small probability.  However,
it is difficult to imagine being rid of EPR pairs as they appear to be
the only kind of states that can be trusted and yet allow efficient
gate testing.

We circumvent the aforementioned difficulty using the fact that our
circuits can have classically controlled feedback that decides which
gates need to be applied based on some measurement results. More
precisely, given a circuit for a unitary transformation $T$ and an
input $x$, we first measure the $B$-side of the (alleged) EPR
states. This yields a classical state $y$ on the $A$-side. Second, we
design a circuit $T_{x,y}$ whose purpose is to flip the corresponding
bits of $y$ in order to get the input $x$, and to apply the initial
circuit for $T$. Third, we run the modified circuit on the state $y$
that was prepared on the $A$-side. Finally, we test that this modified
circuit implemented the correct computation. This includes verifying
the gates and the preparation of all input states $\ket{x'}$---and in
particular the preparation of $\ket x$---obtained by measuring $\ket
\Psi$ on the $B$-side.

\subsection{An Example}\label{example-circuit}
\begin{figure}[h]
\begin{center}
\includegraphics[width=14cm]{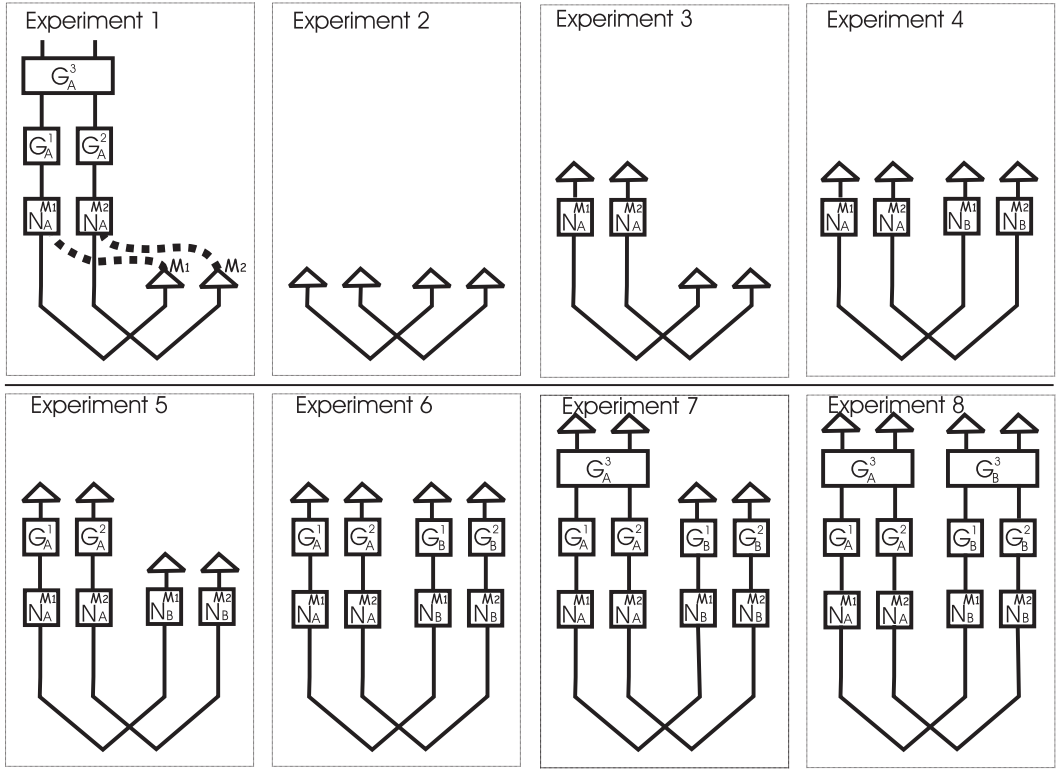}
\end{center}
\caption{The different experiments to test the circuit consisting of
gates $G^3_A G^2_A G^1_A$ on input $\ket{00}$.\label{fig1}}
\end{figure}

As a simple example, in Figure~\ref{fig1} we consider a small 2-qubit
circuit that requires all-zeros as input.

We first run the computation (Experiment 1) once. Suppose the
intermediate measurements on the B-side yield the outcomes $M_1,M_2
\in \{0,1\}$, as indicated in the diagram. The measurement outcomes
determine whether $N^0 = I$ or $N^1 = N$ were applied to the other
halves of the (alleged) EPR pairs, in order to prepare `0' inputs for
the initial circuit we intended to run.

We now wish to check that the output of the circuit is correct. We
carry on implementing Experiments 2 through 8 each a number of times
in $\log(n/\gamma)/\eps^{8}$, where $\eps$ is the required precision
and $\gamma$ is some confidence parameter. In general, the number of
different circuits to be run is linear in $t+n$, where $t$ is the
number of gates in the circuit and $n$ is the number of qubits of the
circuit, so we consider the test to be efficient.

The test circuits correspond to two independent sub-circuits being run
on separate halves of $n$ EPR pairs.  While the gate $G_A^i$ is
purposed to implement the $i$-th step of the circuit, the gate $G_B^i$
should undo $G_A^i$ (by implementing the transpose gate).  There are
two types of tests. The ``conspiracy tests'' (Experiments 2,4,6,8)
verify the effective dimension of the Hilbert spaces to be $2$ for
each computational qubit system at each step of the circuit, and the
``tomography tests'' (Experiments 3,5,7) are characterizing the
unitaries to confirm that they are the correct ones. Since the systems
on each half of the test circuit never interact again, the gates on
each side cannot ``know'' if they are in a conspiracy test, a
tomography test, or the actual computation.

Thus, if all the conspiracy and tomography tests are passed, we are
confident that the actual computation was carried out faithfully, and
any ancillary states are not entangled with the output of the ideal
circuit.

\subsection{The generic Test and its Analysis}
The parameters of our test is a circuit for $T\in\U(2^n)$, that is a
gate decomposition $T^tT^{t-1}\cdots T^1=T$; a binary string
$x\in\{0,1\}^n$; a precision $\eps>0$; and a confidence $\gamma>0$.
We assume that each gate $T^i$ acts on a constant number of qubits
(say $\leq 3)$.  The input is a source of quantum states $\ket{\Psi}$
spread over $n$ pairs of quantum registers; gates $G_A^j$ and $G_B^j$
acting on the same register numbers as $T^j$, for every $j$; auxiliary
gates $N_A^i$ acting on the $i$-th register of $A$; and orthogonal
measurements $(P_{A^i}^a,P_{A^i}^{a+\pi/2})_{a\in\A_0}$ and
$(P_{B^i}^b,P_{B^i}^{b+\pi/2})_{b\in\A_0}$.  The goal is to test that,
firstly, $\sqrt{2^n}\tr_B(P^b_B\ket{\Psi})$ simulates $\ket{b}$ and
that, secondly, the implemented circuit $G_A$ simulates $T$.

{\small\begin{center}
\begin{fmpage}{16cm}
\textbf{Circuit Test
$(T^1,T^2,\ldots,T^t\in\U(2^n),x\in\{0,1\}^n,\eps>0,\gamma>0)$}
\begin{enumerate}
\item Prepare a state $\ket{\Psi}$ of $n$ EPR states into $n$ pairs of
registers $A^1\otimes B^1,\ldots, A^n\otimes B^n$
\item Observe the $B$-side of $\ket{\Psi}$ using the orthogonal
measurement $(P_B^b)_{b\in\{0,\pi/2\}^n}$ and let $y$ be the outcome
\item Let $T_{x,y}$ be the circuit that changes the input $\ket{y}$
into $\ket{x}$ (using some NOT gates), and then applies $T$
\item Prepare on the $A$-side the circuit $G_A$ implementing $T_{x,y}$
with respect to its gate decomposition using the $t$ gates of $G_A^j$
and at most $n$ gates of $N_A^i$. Let $t'\leq t+n$ be the total number of
gates.
\item Run the circuit on the $A$-side and measure the outcome using
the orthogonal measurement $(P_A^a)_{a\in\{0,\pi/2\}^n}$\label{circuit-step}
\item\label{step6}
Approximate all the following statistics by repeating $O(\tfrac{\log
(n/\gamma)}{\eps})$ times the following measurements (where we use the
notation of Theorem~\ref{circuit-theorem}):
\begin{enumerate}
\item Measure $\ket{\Psi}$ with respect to
$(P_{A^i}^{a^i}\otimes P_{B^i}^{b^i})_{a^i,b^i\in \A_0}$, for every
$i=1,2,\ldots,n$.
\item For every $j=1,\ldots,t'$: Measure $\ket{\Psi^j}$
with respect to $(P_{A^i}^{a^i}\otimes
P_{B^i}^{b^i})_{a^i,b^i\in \A}$, for every $i\in I^j$.
\item For every $j=1,\ldots,t'$: Measure $\ket{\Psi'^j}$
with respect to $(P_{A^{I^j}}^{a}\otimes P_{B^{I^j}}^{b})_{a, b \in \A_0^{I^j}}$.
\end{enumerate}
\item Accept if all the statistics are correct up to an additive error $\eps$
\end{enumerate}
\end{fmpage}
\end{center}}

\begin{theorem}
Let $T^1,T^2,\ldots,T^t\in\U(2^n),x\in\{0,1\}^n, \eps>0,\gamma>0$.

If \textbf{Circuit Test$(T^1,T^2,\ldots,T^t,x,\eps,\gamma)$} accepts
then, with probability $1-O(\gamma)$, the outcome probability
distribution of the circuit (in step~\ref{circuit-step}) is at total
variance distance $O((t+n)\eps^{1/8})$ from the distribution that
comes from the measurement of $T^tT^{t-1}\cdots T^1\ket{x}$ by
$(\ketbra{a})_{a\in\{0,\pi/2\}^n}$.

Conversely, if \textbf{Circuit
Test$(T^1,T^2,\ldots,T^t,x,\eps,\gamma)$} rejects then, with
probability $1-O(\gamma)$, at least one of the state $\ket{\Psi}$, the
gates $G_A^i, G_B^i$ and $N_A^i$ is not $O(\eps)$-equivalent to
respectively either $\ket{\Phi_n^+}$, $(\ketbra{a}_{A^i_c})_{a\in\A},
(\ketbra{b}_{B^i_c})_{b\in\A})$, $T^i,{}^t(T^i)$ and
$\mathrm{NOT}_{A^i_c}$, on $S=\Span(P_A^a\otimes P_B^b \ket{\Psi}:
a,b\in\A^n)$ with respect to the projections $(P_A^a\otimes
P_B^b)_{a,b\in\A^n}$.

Moreover \textbf{Circuit Test$(T^1,T^2,\ldots,T^t,x,\eps,\gamma)$}
consists of $O(\tfrac{t n}{\eps}\log(n/\gamma))$ samplings.
\end{theorem}

\begin{proof}
We first describe the use of the hypotheses we made in
Section~\ref{intro} on our testing model.  The assumption (H4) of
trusted classical control is used to ensure that the circuit has the
same behavior on $P_B^y \ket\Psi$ as it would have on $\ket \Psi$.
Hypothesis (H3) implies that we can repeat several times the same
experiment, and hypotheses (H1) and (H2) allow us to state which parts
of our system are separated from the others.

First, using the Chernoff-Hoeffding bound, we know that the
expectation of any bounded random variable can be approximated within
precision $O(\eps)$ with probability $1-O(\gamma)$ by $\tfrac{\log
(1/\gamma)}{\eps^2}$ independent samplings.  Moreover if the
expectation is lower bounded by a constant, then $\tfrac{\log
(1/\gamma)}{\eps}$ independent samplings are enough.  In our case, the
random variable is the two possible outcomes of a measurement.  Call
them $0$ or $1$. Since we can count both $0$ and $1$ outcomes, one of
the corresponding probabilities is necessarily at least $1/2$.
Therefore we get that each statistics we have from \textbf{Circuit
Test} are approximated within precision $O(\eps)$ with probability
$1-O(\gamma)$. 
%%%FRED
From now on, we assume that each statistics 
has been approximated within this precision.

The second part of the theorem is the soundness of \textbf{Circuit
Test}.  We prove it by contraposition.  Namely, if our objects are at
distance at most $\eps$ from ones that exactly satisfies the
statistics, then their own statistics has a bias which is upper
bounded by $O(\eps)$, thanks to the statistics properties of
$\ell_2$-norm on states and the corresponding operator norm.

The rest of the proof now consists in proving 
the first part of the theorem, that is the robustness of \textbf{Circuit
Test}.  We first derive the correct simulation of the implemented
circuit using the approximate version of
Corollary~\ref{circuit-input-corollary}, that we get using
Theorems~\ref{a-MY-theorem} and~\ref{a-nqubit-theorem}.  
More precisely, using Corollary~\ref{a-MY-n-sep} for the initial source
we get that $\ket{\Psi}$ is
$O(n\eps^{1/4})$-equivalent to $\ket{\Phi^+_n}$ on $S$.
For other steps, due to the application of the $j$-th gate,
the state $\ket{\Psi^j}$ is not necessarily a separable state across the $n$-registers.
So we apply Corollary~\ref{a-MY-n} on the registers where the  $j$-th gate is applied,
that is on a constant number of register, which gives the required
$O(\eps^{1/4})$-equivalence on the corresponding registers.
Then Theorem~\ref{a-nqubit-theorem} concludes
that the $j$-th gate is
$O(j\eps^{1/8})$-equivalent to the expected one, similarly for the
intermediate states of the circuit and for the measurements.
Note the error propagation is controlled by two properties: the stability of
the $\ell_2$ operator-norm by tensor product composition, and the
triangle inequality of the norm.  

Now we focus on the run of $T_{x,y}$ in Step~\ref{circuit-step}.
%\label{a-MY-theorem}
%\label{circuit-input-corollary}
First we justify that the
(normalized) outcome state $\sqrt{2^n} P_B^y\ket{\Psi}\in S$ of the
measurement $(P_B^b)_{b\in\{0,\pi/2\}^n}$ is
$O(n\eps^{1/4})$-equivalent to $\ket{y}$ with respect to
$(P_A^a)_{a\in\{0,\pi/2\}^n}$ on $ P_B^y(S)$.
Remind that by assumption the initial state $\ket{\Psi}$ is separable across
the $n$ pairs of registers, namely $\ket{\Psi}=\bigotimes_i \ket{\psi^i}$
with $\ket{\psi^i}\in A^i\otimes B^i$. 
For each pair of registers $A^i\otimes B^i$, using Theorem~\ref{a-MY-theorem} 
we get that $\ket{\psi^i}$ is $O(\eps^{1/4})$-equivalent to $\ket{\phi^+}$
with respect to $(P_{A^i}^{a^i}\otimes P_{B^i}^{b^i})_{a^i,b^i\in\A}$
on $S^i=\Span (P_{A^i}^{a^i}\otimes P_{B^i}^{b^i} \ket{\psi^i} : a^i,b^i\in\A)$.
In particular the projections $P_{A^i}^{a^i}\otimes P_{B^i}^{b^i}$
 are also $O(\eps^{1/4})$-equivalent to $\ketbra{a^i}\otimes\ketbra{b^i}$ on $S^i$.
Therefore the normalized outcome state $\sqrt{2} P_B^{y^i}\ket{\psi^i}$ (which is in $S^i$)
is $O(\eps^{1/4})$-equivalent to $\ket{y^i}$ with respect to
$(P_{A^i}^{a^i})_{a^i\in\{0,\pi/2\}}$ on $P_{B^i}^{y^i}(S^i)$.
We then get our equivalence for the whole outcome state using those intermediate equivalences together with the stability of the $\ell_2$ operator-norm by tensor product composition, and the
triangle inequality of the norm.  
%
%
%
%{From} Corollary~\ref{a-MY-n-sep}, the initial state $\ket{\Psi}$ is
%$O(n\eps^{1/4})$-equivalent to $\ket{\Phi^+_n}$ on $S$, and therefore
%$S$ is $O(n\eps^{1/4})$-equivalent to $\Hi_{2^{2n}}$.  Thus the
%(normalized) outcome state $\sqrt{2^n} P_B^y\ket{\Psi} \in S$ of the
%measurement $(P_B^b)_{b\in\{0,\pi/2\}^n}$ is
%$O(n\eps^{1/4})$-equivalent to $\ket{y}$ with respect to
%$(P_A^a)_{a\in\{0,\pi/2\}^n}$ on $ P_B^y(S)$.  There is no blow up of
%the error since the equivalence is for unit vectors of $S$.  

Lastly, we combine the above approximate equivalences, one for the circuit and one for the input,
and get that the outcome distribution is at total variation distance at most
$O((t+n)\eps^{1/8})$ from the expected one.
\end{proof}

%\newpage

\bibliographystyle{alpha} \bibliography{qtest}

%\newpage

\begin{appendix}

\section{A Conspiracy for the Hadamard Test of~\cite{dmms00}}\label{example}
This example is due to Wim~van~Dam. Consider the test for the Hadamard
gate of~\cite{dmms00}. It essentially consisted of verifying that
starting with $\ket{0}$ or $\ket{1}$ followed by a Hadamard gate and a
measurement resulted in a $50\% - 50\%$ distribution of `0' and `1'
outcomes, and starting with $\ket{0}$ followed by two Hadamard gates
and a measurement resulted in a `0' outcome $100\%$ of the time.

A very simple conspiracy (i.e., alternative explanation of the gate
action that is not equivalent to the claimed action) that foils this
test is the following. The qubit system is actually a $4$-state system
consisting of two qubits. Our alleged $\ket{0}$ state corresponds to
$\ket{00}$, and our alleged $\ket{1}$ state corresponds to
$\ket{11}$. The alleged Hadamard gate simply maps $\ket{00} \mapsto
\ket{01}$, $\ket{01} \mapsto \ket{00}$, $\ket{11} \mapsto \ket{10}$,
$\ket{10} \mapsto \ket{11}$. In other words, our alleged $\ket{0} +
\ket{1}$ actually corresponds to $\ket{01}$ and our alleged $\ket{0} -
\ket{1}$ actually corresponds to $\ket{11}$. Our measurement operation
simply outputs one of the two bits at random.  So measuring $\ket{00}$
always results in `0', measuring $\ket{11}$ always results in '1', and
measuring $\ket{01}$ or $\ket{10}$ results in `0' or `1' each with
probability $\frac{1}{2}$.

Note that this physical system would pass the test of~\cite{dmms00}
for the Hadamard gate, but clearly the system is not implementing the
Hadamard gate. For example, if this apparatus were being used to
implement the quantum key distribution of~\cite{bb84}, the result
would be disastrous since a competent eavesdropper could reliably
distinguish all four states. It is imperative for truly secure the
quantum key distribution of~\cite{bb84} that the qubits Alice sends
are truly residing in a 2-dimensional Hilbert space with no crucial
information leaked in extra degrees of freedom or ``side channels''.

\section{EPR test and its Robustness}\label{my}
In this section we sketch the proof of Theorem~\ref{MY-theorem}, so
that we can justify its robustness.

The proof proceeds with two main steps. The first step proves the
existence of a strong equivalence, a notion we define next. The second
shows that this strong equivalence implies the required tensor
equivalence.

\subsection{Strong equivalence}
Intuitively the strong equivalence states the existence of an isometry
between the ideal system and the physical system. Contrarily to the
previously defined notion of equivalence, this does not require the
use of an auxiliary system (see Definition~\ref{def_eq}).

\begin{definition}
Let $S$ be an $N$-dimensional subspace of an Hilbert space $H$, and
$U\in\Iso(S, \Hi_N)$. We say that $S$ is {\em strongly $U$-equivalent}
to $\Hi_N$ (with respect to $(P^w)_{w\in\mathcal{W}}$) if $S$ is 
$P^w$-invariant (that is $P^w(S)\subseteq S$) and $P^w =_{S}
U^\+\ketbra{w} U$, for every $w$.
\end{definition}

The above definition is equivalent to say that the following diagram
is commutative:
$$
\begin{array}{rcl}
S &\xrightarrow{P^w}& S \\ 
U\downarrow & & \uparrow U^\+\\
\Hi_N&\xrightarrow{\ketbra{w}} & \Hi_N 
\end{array}
$$

Now, we can define the strong equivalence between two states and
between two unitary transformations.
\begin{definition}
Let $S$ be a subspace of $H$.
A state $\ket{\psi}\in S$ is  {\em strongly $U$-equivalent}  to 
a state $\ket{\phi}\in\Hi_N$  on $S$
(with respect to $(P^w)_{w\in\mathcal{W}}$), if 
\vspace*{-7pt}\begin{enumerate}
\item $S$ is strongly $U$-equivalent to $\Hi_N$
\item $\ket{\psi} =  U^\+ \ket{\phi}$.
\end{enumerate}
\end{definition}
\begin{definition}
Let $S$ be a subspace of $H$.
A unitary transformation $G\in\U(H)$
is {\em strongly $(U,V)$-equivalent} 
to a unitary transformation $T\in\U(\Hi_N)$ on $S$
(with respect to $(P^w)_{w\in\mathcal{W}}$), if
\vspace*{-7pt}\begin{enumerate}
\item $S$ is strongly $U$-equivalent to $\Hi_N$,
\item $S'=G(S)$ is $V$-equivalent to $\Hi_N$,
\item $G =_{S} V^\+  T  U$.
\end{enumerate}
\end{definition}

\subsection{Outline of the proof of Theorem~\ref{MY-theorem}}\label{my3}
This theorem is essentially obtained by proving two intermediate
lemmas. The first one states that $\ket\psi$ is strongly equivalent to
$\ket{\phi^+}$ without reference to the tensor product structure of $H
= A\otimes B$. The second one recovers this structure and ends the
proof.

\begin{lemma}\label{MY-lemma1}
Let $S=\Span((P_A^a\otimes P_B^b\otimes \id_C) \ket{\psi} :
a,b\in\A)$.  Under the hypothesis of Theorem~\ref{MY-theorem}, there
exists an isometry $U\in\Iso(S,\Hi_4)$ such that $\ket{\psi}$ is
strongly $U$-equivalent to $\ket{\phi^+}$ on $S$.
\end{lemma}

This result is obtained in three steps. First the state $\ket{\psi}$
satisfies the main property of any EPR state: the outcome state does
not depend on which side the measurement is performed.
\begin{proposition}[{\cite[Prop. 1]{my03}}]\label{MY-prop1}
$P_A^a \ket{\psi}=P_B^a \ket{\psi}=(P_A^a\otimes P_B^a\ket\otimes
\id_C){\psi}$, for every $a\in \A$.
\end{proposition}
Second, the statistical behavior of the measurement outcomes is
rewritten in terms of geometric properties of the collapsed states.
For $\alpha\neq\beta\in \A_0$, define $\Theta_{\alpha,\beta} = \{(a,b)
: a=\alpha,\alpha+\pi/2, b=\beta,\beta+\pi/2\}$, and $B_{\alpha,\beta}
= ((P_A^a \otimes P_B^b \otimes \id_C)\ket{\psi}:
(a,b)\in\Theta_{\alpha,\beta})$.
\begin{proposition}[{\cite[Prop. 2]{my03}}]\label{MY-prop2}
Let $\alpha\neq\beta\in \A_0$.  The four vectors of $B_{\alpha,\beta}$
are mutually orthogonal and have the same length as the corresponding
ideal vectors $((\ketbra{a}\otimes\ketbra{b})\ket{\phi^+} :
a,b\in\Theta_{\alpha,\beta})$.
\end{proposition}
These geometric properties for any $\alpha\neq\beta$ can be rewritten
under the strong-equivalence notion. That is
$S_{\alpha,\beta}=\Span(B_{\alpha,\beta})$ is strongly
$U_{\alpha,\beta}$-equivalent to $\Hi_4$, where $U_{\alpha,\beta}$ is
the isometry that maps $P_A^a P_B^b \ket{\psi}$ to
$(\ketbra{a}\otimes\ketbra{b})\ket{\phi^+}$ for $a,b \in
\Theta_{\alpha, \beta}$.

The third step states that in fact $S_{\alpha,\beta} =
S_{\alpha',\beta'}=S$, for every $\alpha'\neq\beta'$, and that
$U_{\alpha, \beta}$ is independent from the choice of $\alpha,\beta$.
\begin{proposition}[{\cite[Prop. 3]{my03}}]\label{MY-prop3}
Let $\alpha,\beta,\alpha',\beta'\in\A_0$ be such that
$\alpha\neq\beta$ and $\alpha'\neq\beta'$.  The vectors of
$B_{\alpha,\beta}$ are in the real span of
$B_{\alpha',\beta'}$. Moreover, the matrix corresponding to the basis
change from $B_{\alpha,\beta}$ to $B_{\alpha',\beta'}$ is identical to
the one of the ideal case.
\end{proposition}
Lemma~\ref{MY-lemma1} follows directly from this last observation.

The next lemma ends the proof of Theorem~\ref{MY-theorem}. It shows
that the strong equivalence, which involves a global isometry $U$,
implies the tensor equivalence over $S$. That is, it involves only
local unitary transformations over $\bar A$ and $\bar B$. Moreover the
subspace $S$ where the tensor equivalence holds can be extended to the
tensor product of the supports of $\ket{\psi}$ on the $A$-side and on
the $B$-side.

\begin{lemma}\label{MY-lemma2}
Let $S=\Span((P_A^a\otimes P_B^b\otimes \id_C) \ket{\psi} :
a,b\in\A)$.  Assume that $\ket{\psi}$ is strongly $U$-equivalent
$\ket{\phi^+}$ on $S$, then there exist two unitary transformations
$U_{\bar A}\in\U(\bar A)$ and $U_{\bar B}\in\U(\bar B)$ such that
$\ket{\psi}$ is $(U_{\bar A}\otimes U_{\bar B})$-equivalent to
$\ket{\phi^+}$ on $S$.
\end{lemma}

The proof of this lemma is constructive, that is the transformations
$U_{\bar A}$ and $U_{\bar B}$ will be defined explicitly in terms of
the projectors $P_A^a$ and $P_B^b$. More precisely, we define the NOT
and control-NOT using the given orthogonal projections.  The
transformations $U_{\bar A}$ and $U_{\bar B}$ are constructed using
the decomposition of a SWAP gate as two control-NOT gates (the
logical qubit being in the $\ket 0$ state as required by the
embedding $\inj_A$ of $A$ in $\bar A$). It is then checked that
$U_{\bar A}$ and $U_{\bar B}$ fulfill the conclusions of the lemma.

First, the NOT gate on $A$ is defined by $N_A = 2 P_A^{\pi/4} -
\id_A$.  The NOT gate on $A_c$ is denoted by $N_{A_c}$.  Then, the
c-NOT gates on $\bar A$ are defined by $c_{A_c}\text{-}N_A=\ketbra 0
\otimes \id_A + \ketbra{\pi/2} \otimes N_A$, and
$c_A\text{-}N_{A_c}=\id_{A_c} \otimes P_A^0 + N_{A_c} \otimes
P_A^{\pi/2}$.  Finally, the transformation $U_{\bar A}$ which extracts
the state of the physical qubit included in $A$ by swapping it into
$A_c$ is given by $U_{\bar A}= (c_{A_c}\text{-}N_A)
(c_A\text{-}N_{A_c})$.

The first observation is that all these transformations are
necessarily unitary since they involve projections that come from
orthogonal measurements. Moreover, they are all equivalent to their
ideal mapping on $S$, namely to the transformations $N_2$,
$c_{A_c}\text{-}N_2$ and $c_2\text{-}N_{A_c}$, which are defined by
substituting $A$ with $\Hi_2$. In the rest of this section we are
using simultaneously many different spaces, hence we explicitly write
the appropriate injection $\inj_A$ and projection $\proj_A$.
\begin{proposition}[{\cite[Eq. 10]{my03}}]\label{MY-prop10}
Let $I\in\Iso(A_c, \Hi_2)$ be the canonical isometry between $A_c$ and
$\Hi_2$.
\vspace*{-7pt}\begin{enumerate}
\item $N_A\otimes\id_B$ is strongly $(U,U)$-equivalent to $N_2\otimes
\id_2$ on $S$.
\item $c_{A_c}\text{-}N_A\otimes\id_B$ is strongly $(I\otimes
U,I\otimes U)$-equivalent to $c_{A_c}\text{-}N_2\otimes \id_2$ on
$\inj_A(S)=\ket{0}_{A_c}\otimes S$.
\item $c_A\text{-}N_{A_c}\otimes\id_B$ is strongly $(I\otimes
U,I\otimes U)$-equivalent to $c_2\text{-}N_{A_c}\otimes\id_2$ on
$\inj_A(S)=\ket{0}_{A_c}\otimes S$.
\end{enumerate}
\end{proposition}
Therefore $U_{\bar A}$ is strongly $(I\otimes U,I\otimes
U)$-equivalent to the SWAP gate between $\Hi_2$ and $A_c$ on
$\ket{0}_{A_c}\otimes S$. The transformation $U_{\bar B}$ can be similarly defined on
$\bar B$ with the same above properties.

To conclude the proof, it is sufficient to check that the tensor
product equivalence holds on $S$. Using the above properties of
$U_{\bar A}$ and $U_{\bar B}$, this requires only a bit of algebra.
In effect, we get \cite[Eq. 11 \&12]{my03}:
\begin{equation}\label{MY-eq10}
\begin{split}
\ket{0}_{A_c}\otimes\ket{0}_{B_c}\otimes\ket{\psi} &=(U_{\bar
A}^\+\otimes U_{\bar B}^\+) (\ket{\phi^+}_{A_c B_c}\otimes (U^\+
\ket{00}_{\Hi_4})_{AB}),\\
\forall \ket{\varphi}\in S,\quad (P_A\otimes
\id_B)\ket{\varphi}&=\proj_A (U_{\bar A}^\+\otimes\id_B)
(\ketbra{a}{a}_{A_c} \otimes \id_{AB}) (U_{\bar A}\otimes\id_B)
\inj_A(\ket{\varphi}),\\
\forall \ket{\varphi}\in S,\quad (\id_A\otimes
P^b_B)\ket{\varphi}&=\proj_B (\id_A\otimes U_{\bar B}^\+)
(\ketbra{b}{b}_{B_c} \otimes \id_{AB}) (\id_A \otimes U_{\bar B})
\inj_B(\ket{\varphi}).
\end{split}
\end{equation}

These equations can be summarized in the following proposition.
\begin{proposition}[{\cite[Eq. 11 \&12]{my03}}]\label{MY-prop11}
$\ket{\psi}$ is $(U_{\bar A}\otimes U_{\bar B})$-equivalent to
$\ket{\phi^+}$ on $S$.
\end{proposition}

The tensor-equivalence can then be extended to the tensor product of
the respective supports using~\cite[Prop. 4]{my03}.

This ends the summary of the proof of Mayers and Yao's result.

\subsection{Robustness}
The notion of strong equivalence is extended into $\eps$-strong
equivalence in the same way equivalence was extended into
$\eps$-equivalence. In particular, for the $\eps$-strong equivalence,
the subspace $S$ does not need to be $P^w$-invariant anymore. However,
we require that each unit vector of $P^w(S)$ is at distance at most
$\eps$ from a vector of $S$.

The proof of Theorem~\ref{a-MY-theorem} is again in two steps
by making Lemmas~\ref{MY-lemma1}\&\ref{MY-lemma2} robust.
One way of stating an approximated equivalence is to derive it from an
orthogonal basis using the following proposition.
\begin{proposition}\label{approx-prop}
Let $B$ be a finite set of orthogonal and unit vectors of $H$. Define
$S=\Span (B)$.  Let $M$ and $N$ be two linear transformations on $H$
such that $M\ket\psi =^\eps N\ket{\psi}$, for every $\ket{\psi}\in B$.
Then $\norm{M-N}_S\leq\sqrt{\size{B}}\eps$.
\end{proposition}

Below, $S$, $S_{\alpha, \beta}$ and $S_0$ are defined as in the
previous section.
\begin{lemma}\label{a-MY-lemma1}
Under the hypothesis of Theorem~\ref{a-MY-theorem}, there exists an
isometry $U\in\Iso(S_0,\Hi_4)$ such that $\ket{\psi}$ is strongly
$(O({\eps}^{1/4}),U)$-equivalent to $\ket{\phi^+}$ on $S$.
\end{lemma}

\begin{proof}[Sketch of proof]
We follow the structure of the proof of Lemma~\ref{MY-lemma1}.  Let
$\delta=\sqrt{\eps}$.  First we rephrase Propositions~\ref{MY-prop1}
and~\ref{MY-prop2} easily since they directly derive from the
statistics of $\ket{\psi}$.  This give us $P_A^a \ket{\psi}=^{\delta}
P_A^a\otimes P_B^a\ket{\psi}$ and $P_B^a \ket{\psi}=^{\delta}
P_A^a\otimes P_B^a\ket{\psi}$, for every $a\in A$. Moreover, for every
$\alpha\neq\beta$, the four states of $B_{\alpha,\beta}$ are still
orthogonal but their lengths are now approximately correct up to an
additive error $\delta$.

Since the sets $S_{\alpha,\beta}$ will not necessarily coincide with
each other anymore, we first fix arbitrarily
$B_0=B_{\alpha_0,\beta_0}$, for some $\alpha_0\neq\beta_0$, and
$S_0=S_{\alpha_0,\beta_0}$.  Then we will show that any vector from
$S$ is close to a vector of $S_0$.

Following the proof of Proposition~\ref{MY-prop3}, which is based on
geometrical arguments in dimension 8, one can prove that the
re-normalized vectors of $B_{\alpha,\beta}$ are now at distance at
most $\sqrt{\delta}$ from a vector of the real span of
$B_{0}$. Moreover, the basis change matrix between $B_{\alpha,\beta}$
and $B_{0}$ corresponds to the one of the ideal case up to an additive
error in $O(\sqrt{\delta})$.

We now construct $U$ in a way similar to that of the perfect
case. Because the length of the four vectors in $B_0$ is not
necessarily correct, $U$ is defined after re-normalizing them. In
short, the isometry $U$ is the isometry that maps the (re-normalized)
states $P_A^a P_B^b \ket{\psi}$ to (re-normalized)
$(\ketbra{a}\otimes\ketbra{b})\ket{\phi^+}$ for every
$a,b\in\Theta_{\alpha_0,\beta_0}$.

The conclusions of the Lemma hold on $S_0$ from
Proposition~\ref{approx-prop}.

A consequence is that for every $\alpha_0\neq\beta_0\in\A_0$ and
$\alpha\neq\beta\in\A_0$, the spaces $S_0=S_{\alpha_0,\beta_0}$ and
$S_{\alpha,\beta}$ are close. For unit vectors $\ket{\psi_0}\in S_{0}$
and $\ket{\psi}\in S_{\alpha,\beta}$ we have $\max_{\ket{\psi_0}}
\min_{\ket{\psi} } \norm{\ket{\psi_0}-\ket{\psi}} \in
O({\eps}^{1/4})$.  This justifies that the conclusion can be extended
from $S_0$ to $S$ with an additional error term in $O({\eps}^{1/4})$.
\end{proof}

\begin{lemma}\label{a-MY-lemma2}
Assume that $\ket{\psi}$ is strongly $(\eps,U)$-equivalent to
$\ket{\phi^+}$ on $S$, then there exist two unitary transformations
$U_{\bar A}\in\U(\bar A)$ and $U_{\bar B}\in\U(\bar B)$ such that
$\ket{\psi}$ is $(O(\eps),U_{\bar A}\otimes U_{\bar B})$-equivalent
to $\ket{\phi^+}$ on $S$.
\end{lemma}

\begin{proof}[Sketch of proof]
We again follow the structure of the proof of Lemma~\ref{MY-lemma2}.
Define $U_{\bar A}$ and $U_{\bar B}$ in the very same way.  These are
still unitary transformations even if the statistics are not exact.
The first modifications start with Proposition~\ref{MY-prop10}, where
an additive error term $2\eps$ comes from the use of two projections
in each expression of $N_A$, $c_{A_c}\text{-}N_A$ and
$c_2\text{-}N_{A_c}$, such that these projections are all
$\delta$-equivalent to their ideal projections.
\vspace*{-7pt}\begin{enumerate}
\item $N_A\otimes\id_B$ is strongly $(2\eps,U,U)$-equivalent to
$N_2\otimes \id_2$ on $S_0$.
\item $c_{A_c}\text{-}N_A\otimes\id_B$ is strongly $(2\eps,I\otimes
U,I\otimes U)$-equivalent to $c_{A_c}\text{-}N_2\otimes \id_2$ on
$\ket{0}_{A_c}\otimes S_0$.
\item $c_A\text{-}N_{A_c}\otimes\id_B$ is strongly $(2\eps,I\otimes
U,I\otimes U)$-equivalent to $c_2\text{-}N_{A_c}\otimes\id_2$ on
$\ket{0}_{A_c}\otimes S_0$.
\end{enumerate}

Then Equations~\eqref{MY-eq10} are also extended up to an additive
error term in $O(\eps)$, which ends the sketch of the proof.
\end{proof}

Note that our robust statements can only be made on $S$. Any results
that have been extended to the support of $\ket{\psi}$ on the $A$-side
using~\cite[Prop. 4]{my03} cannot be made robust, at least
independently of the dimension of $A_{\ket{\psi}}$, because of the
instability of~\cite[Prop. 4]{my03}.

\subsection{Proof of corollary~\ref{a-MY-n}}
We proceed by induction over $n$. From theorem~\ref{a-MY-theorem}, we
have that $\ket \Psi$ is $\eps^{1/4}$-equivalent to $\ket{\phi^+}$ on
$\Span\{P_{A^n}^{a^n} P_{B^n}^{b^n} \ket\Psi: a_n, b_n \in \A\}$ with
respect to the measurements $P_{A^n}^{a^n} P_{B^n}^{b^n}$. Fix $a_n
\in \{0, \tfrac{\pi}{2}\}$ and $b_n \in \{\tfrac{\pi}{4},
\tfrac{\pi}{4} + \tfrac{\pi}{2}\}$. Then, the state $P_{A^n}^{a^n}
P_{B^n}^{b^n} \ket\Psi / \norm{P_{A^n}^{a^n} P_{B^n}^{b^n} \ket\Psi}$
is $2\eps$-simulating $\ket{\phi^+}$ with respect to $P_{A^i}^{a^i}
P_{B^i}^{b^i}$ for every $1\leq i\leq n-1$. Applying our hypothesis
for $n-1$, we get that $P_{A^n}^{a^n} P_{B^n}^{b^n} \ket\Psi /
\norm{P_{A^n}^{a^n} P_{B^n}^{b^n} \ket\Psi}$ is $2\times
4^{n-1}\eps^{1/4}$-equivalent to $\ket{\Phi^+_{n-1}}$ on
$\Span\left\{\left(\bigotimes_{i=1}^{n-1} (P_{A^i}^{a^i}
P_{B^i}^{b^i})\right) (P_{A^n}^{a^n} P_{B^n}^{b^n})\ket\Psi : a^i, b^i
\in \A, \ 1\leq i\leq n-1\right\}$. Note that the unitaries that are
constructed for obtaining this equivalence are built independently
from the value of $a^n$ and $b^n$. Therefore, using
Proposition~\ref{approx-prop}, we obtain that $S = \Span \{P_A^a P_B^b
\ket \Psi : a,b \in \A^n\ \}$ is $4^n\eps^{1/4}$-equivalent to $A_c
\otimes B_c$ with respect to $\bigotimes_{i=1}^{n-1} P_{A^i}^{a^i}
P_{B^i}^{b^i}$. Since it would have been possible to single out say
$A^1\otimes B^1$ instead, combining the two results gives $S = \Span
\{P_A^a P_B^b \ket \Psi : a,b \in \A^n\ \}$ is
$4^n\eps^{1/4}$-equivalent to $A_c \otimes B_c$ with respect to $P_A^a
P_B^b$.

The fact that $\ket \Psi =^{4^n \eps^{1/4}} U_{\bar A} \otimes U_{\bar
B} \ket{\Phi^+_n} \ket\chi$ can be derived from
Theorem~\ref{a-MY-theorem} applied to each $A^i \otimes B^i$ pair
independently.

\subsection{Proof of corollary~\ref{a-MY-n-sep}}
This corollary follows directly from Theorem~\ref{a-MY-theorem}
applied to each $A^i\otimes B^i$ independently when one recognizes
that the separability condition implies that $\ket\Psi =
\left(\bigotimes_i \tr_{ABC - A^iB^i} \ket\Psi\right) \otimes
\tr_{AB}\ket\Psi.$

\section{Proof of Theorem~\ref{a-nqubit-theorem}}\label{a-tomography}

The structure of the proof follows the one presented for testing
$1$-qubit real gates when probabilities are perfectly known.

The $\eps$-simulation of $(T\otimes\id_{2^n}) \ket{\Phi^+_n}$ by
$G_A\ket\Psi$ can be rewritten within the density matrix formalism as
\begin{equation*}
\tr\left((P_A^a\otimes P_B^b\otimes \id_C)G_A \ketbra\Psi
G_A^\+\right) =^{\eps} \tr \left( (\ketbra a \otimes \ketbra b
)(T\otimes \id_{2^n})\ketbra{\Phi^+_n} (T^\+\otimes
\id_{2^n})\right),
\end{equation*}
for any $ a, b \in \A^n$. Here, $\ketbra a$ is a shorthand notation
for $\bigotimes_{i=1}^n \ketbra{a^i}$. Using that $(\id_{2^n}\otimes
\ketbra{b})\ketbra{\Phi^+_n} = \tfrac{1}{2^n} \ketbra{b}\otimes
\ketbra{b}$ and the commutativity of the trace operator, we get
\begin{equation*}
\tr \left((G_A^\+ P_A^a G_A \otimes P_B^b)\ketbra\Psi\right) =^{\eps}
\frac{1}{2^n}\tr\left(T^\+ \ketbra a T \ketbra b \right).
\end{equation*}
Since $\ket\Psi$ is $\eps$-equivalent to $\ket{\Phi^+_n}$, we have
\begin{equation}\label{eq:above}
\tr\left(R^a_{\bar A\bar B C} (\ketbra{b}_{A_c} \otimes
\ketbra{b}_{B_c} \otimes \ketbra{\chi}_{ABC} \right) =^{O(\eps)}
\tr\left(T^\+ \ketbra a T \ketbra b \right),
\end{equation}
where $R^a_{\bar A\bar B C}$ is a positive semi-definite operator
$R^a_{\bar A\bar B C} = (U_{\bar A}\otimes U_{\bar B}\otimes \id_C)
G_A^\+ P_A^a G_A (U_{\bar A}^\+\otimes U_{\bar B}^\+\otimes \id_C)$ on
$\bar A \otimes \bar B \otimes C$. Above, the vector $\ket\chi_{ABC}$
is given by the tensor $\eps$-equivalence of $\ket\Psi$ to
$\ket{\Phi^+_n}$. Equation~\ref{eq:above} can easily yield the
equations required to apply Lemma~\ref{lem:tomo_N} for performing the
tomography of $R^a_{\bar A\bar B}$. For $b,b' \in \A$
\begin{equation*}
\tr\left(R^a_{\bar A\bar B C} (\ketbra b_{A_c} \otimes \ketbra{b'}_{B_c}
\otimes \ketbra\chi_{ABC}) \right) =^{O(\eps)} \tr\left(T^\+ \ketbra a
T \ketbra b \right).
\end{equation*}
Now Lemma~\ref{lem:tomo_N} can be applied to ${}_{ABC}\bra\chi
{}_{B_c}\bra{b'} R^a_{\bar A\bar B C}\ket{b'}_{B_c}\ket\chi_{ABC}$ for
any $b' \in \A$ and its conclusion rewritten as
\begin{equation*}
R^a_{\bar A \bar B C} =^{2^{O(n)}\sqrt{\eps}}_{A_c\otimes B_c\otimes
{\ket{\chi}_{ABC}}} (T^\+ \ketbra a T)\otimes \id_{A \otimes {\bar B}
\otimes C}.
\end{equation*}

The $\eps$-tensor equivalence of $G_A(S)$ with $A_c \otimes B_c$ also
gives (removing obvious identities):
$$P_A^a =_{G_A(S)}^{\eps} (V_A^\+ \otimes U_B^\+) \ketbra a_{A_c}
(V_A \otimes U_B).$$ Using this equality we obtain
\begin{equation*}
\ketbra a_{A_c} (V_{\bar A} \otimes U_{\bar B} ) G_A (V_{\bar A}^\+
\otimes U_{\bar B}^\+ )T^\+ =_{A_c \otimes B_c \otimes
\ket{\chi}_{ABC}}^{2^{O(n)}\sqrt{\eps}} T^\+ (\ketbra a_{A_c} (V_{\bar A}
\otimes U_{\bar B}) G_A (V_{\bar A}^\+ \otimes U_{\bar B}^\+).
\end{equation*}
Lemma~\ref{lem:commut} concludes that:
\begin{equation*}
G_A =_S^{2^{O(n)}\sqrt{\eps}} (V_{\bar A}^\+ \otimes U_{\bar B}^\+ \otimes
\id_C) (T \otimes W \otimes \id_{\bar B\otimes C}) (V_{\bar A} \otimes
U_{\bar B} \otimes \id_C),
\end{equation*}
which ends the proof.

\section{Technical lemmas for Exact and Approximate Tomography}

\begin{lemma}\label{lem:tomo_N}
Let $n \geq 1$ and  $H = \Hi_2^{\otimes n}$.
% For each factor of $H$ define the computational basis $\{\ket 0 ,
%\ket {\tfrac{\pi}{2}}\}$ and the state $\ket{\tfrac{\pi}{4}} =
%\tfrac{1}{\sqrt 2}(\ket 0 + \ket{\tfrac{\pi}{2}})$. 
Let $\ket\gamma$
be a unit vector of $H$ belonging to the real span of the states
$\bigotimes_{i=1}^n \ket{b_i }$ for $(b_i)_i \in
\{0,\tfrac{\pi}{2}\}^n$. Let $\rho$ be a positive semi-definite matrix
over $H$ such that
\begin{equation}\label{eq:tomo}
\forall (b_i)_i \in \{0,\tfrac{\pi}{4},\tfrac{\pi}{2}\}^n, \
\tr\left(\rho \bigotimes_{i=1}^n \ketbra{b_i}\right) =^{\eps}
\tr\left(\ketbra{\gamma}\bigotimes_{i=1}^n \ketbra{b_i}\right).
\end{equation}
Then $\rho =^{2^{O(n)}\sqrt{\eps}} \ketbra\gamma$.
\end{lemma}

\begin{proof}
Define the Pauli matrices for each factor of $H$ as $I = \ketbra{0} +
\ketbra{\tfrac{\pi}{2}}=\id_2$, $X = 2 \ketbra{\tfrac{\pi}{4}} - I$, $Z
= \ketbra 0 - \ketbra{\tfrac{\pi}{2}}$ and $Y = \imath Z X$.  
Recall that $X$, $Y$, and $Z$ have trace $1$, their square is $I$,
and they anti-commute.
A property of
the $n$-fold tensor products of the Pauli matrices,
i.e. $\{I,X,Y,Z\}^{\otimes n}$, is to be an (unnormalized) orthogonal
basis for the hermitian matrices over $H$ for the matrix inner product
$(M,N) = \tr M^\+ N$. Note also that the $n$-fold tensor products of
$I,X,Z$ generate all real symmetric matrices of $H$ by linear
combination.

That is we can write
\begin{equation*}
\ketbra\gamma = \sum_{P\in \{I,X,Z\}^{\otimes n}} c(P) P \quad
\mbox{and} \quad \rho = \sum_{P \in \{I,X,Y,Z\}^{\otimes n}} r(P)P,
\end{equation*}
with $c(P) = \tfrac{1}{2^n}\tr(P\ketbra\gamma) \in \mathbb R$ and
$r(P) = \tfrac{1}{2^n}\tr(P\rho) \in \mathbb{R}$. Since any $P \in
\{I,X,Z\}^{\otimes n}$ is a linear combination of the projectors
$\bigotimes_i \ketbra{b_i}$ with $(b_i) \in
\{0,\tfrac{\pi}{4},\tfrac{\pi}{2}\}^n$, using the linearity of the
trace, Equation~\ref{eq:tomo} implies
\begin{equation*}
\forall P \in \{I,X,Z\}^{\otimes n}, \ r(P) =^{2^{O(n)}\eps} c(P).
\end{equation*}

Because $\rho$ is a positive semi-definite matrix, we have
$\tr(\rho^2) \leq \tr(\rho)^2$. 
Using the properties of Pauli matrices,
the left hand side can be rewritten as $\tr(\rho^2)=\sum_P r(P)^2$,
and the right hand side as $r(I^{\otimes n})^2$,
leading to $\sum_P r(P)^2 \leq r(I^{\otimes n})^2$.

Since $\ketbra{\rho}$ is a projector of rank $1$, it satisfies
$\sum_P c(P)^2 = c(I^{\otimes n})^2$.
Since this sum is only over $\{I,X,Z\}^{\otimes n}$, and that
the coefficients $c(P)$ are close to the coefficients $r(P)$,
we obtain $\sum_P r(P)^2 = 2^{O(n)}\eps$, when the sum is taken
over $P\in \{I,X,Y,Z\}^{\otimes n} -\{I,X,Z\}^{\otimes n}$. 

Using that
$\norm{\rho - \ketbra{\chi}} \leq \sum_P \norm{(r(P) - c(P))P}$, and
the fact that $\norm{(r(P) - c(P))P} = |r(P) - c(P)|$, we obtain that
$\rho =^{2^{O(n)}\sqrt{\eps}} \ketbra \gamma$, which ends the proof.
\end{proof}

\begin{lemma}\label{lem:commut}
Let $n\geq 1$ and
$H_1=H_2=\Hi_2^{\otimes n}$. 
%Let
%$(M_A^a)_a$ be a complete set of symmetric operators with real
%coefficients acting on $A_c$ (i.e., any real symmetric matrix on $A_c$ can
%be written as a linear combination of $M_{A_c}^a$'s). 
Let $U\in\U(H_1\otimes H_2)$. 
If for every $a\in\{0,\pi/4,\pi/2\}^n$
the transformation $U$ satisfies $U (\ketbra{a}{a}_{H_1} \otimes
\id_{H_2}) =^{\eps} (\ketbra{a}{a}_{H_1} \otimes
\id_{H_2}) U$, then there exists $W\in\U(H_2)$
such that $U =^{2^{O(n)} \eps} \id_{H_1} \otimes W$.
\end{lemma}

\begin{proof}
The proof uses the fact that any real symmetric matrix on $H_1$ can be written
as a linear combination of $(\ketbra{a}{a}_{H_1})_{a\in\{0,\pi/4,\pi/2\}^n}$.
Since $(\ket{a})_{a\in\{0,\pi/2\}}$ is the computational basis of $H_1$, we can
write $U =
\sum_{i,j\in\{0,\pi/2\}^n} \ketbraa i j \otimes W_{ij}$ for some $W_{ij}$ acting on
$H_2$. By assumption for every $i,j$,
\begin{align*}
U((\ketbraa i j + \ketbraa j i)\otimes \id_{H_2}) & =^{2^{O(n)}\eps} ((\ketbraa
i j + \ketbraa j i)\otimes \id_{H_2}) U ,\\ 
\text{and}\quad\sum_k (\ketbraa k j \otimes
W_{ki} + \ketbraa k i \otimes W_{kj}) & =^{2^{O(n)}\eps} \sum_k (\ketbraa
i k \otimes W_{jk} + \ketbraa j k \otimes W_{ik}),
\end{align*}
which implies 
$\norm{\sum_{i\neq j}W_{ij}} ={2^{O(n)}\eps}$.

Define the operator $W'=\sum_i W_{ii}$.
Then $W'$ satisfies the required conditions, except that $W'$ is not necessarily
in $\U(H_2)$. Since we assumed that $U$ is a unitary transformation,
one can use a Gram-Schmidt  orthonormalization of $W'$ which will
gives a $W''$ which is at distance at most $2^{O(n)}\eps$ from $W'$.
This concludes the proof.
\end{proof}

\end{appendix}

\end{document}